# Unfolding force definition and the unified model for the mean unfolding force dependence on the loading rate

Rafayel Petrosyan[*]

Department of Physics, University of Alberta, Edmonton AB, T6G 2E1, Canada

## ABSTRACT

In single-molecule force spectroscopy experiments, the dependence of the mean unfolding force on the loading rate is used for obtaining information about the energetic and dynamic properties of the system under study. However, it is crucial to understand that different dynamic force spectroscopy (DFS) models are applicable in different regimes, and that different definitions of the unfolding force might be used in those models. Here, for the first time, we discuss three definitions of the unfolding force. We carried out Brownian dynamics simulations in order to demonstrate the difference between these definitions and compare DFS models. Importantly, we derive the dependence of the mean unfolding force for the whole range of the loading rates by unifying three previously reported DFS models. Among the currently available models, this unified model shows the best agreement with the simulated data.

[*] petrosyan@ualberta.ca
https://orcid.org/0000-0003-1888-0379

# INTRODUCTION

In atomic force microscopy (AFM) based dynamic force spectroscopy (DFS) experiments, the unfolding forces are being measured for a wide range of the loading rates. Afterwards, the average unfolding force dependence on the loading rate from the appropriate DFS model is fitted to the corresponding data in order to gain information about the kinetic and energetic properties of the system[1–4]. For instance, the system could be a protein or a ligand-receptor bond, but the theory is general and not limited to these examples. Hereafter, we will be using the proteins' terminology, such as the native and unfolded states.

The first model that was proposed for analyzing DFS data was the phenomenological model widely referred to as the Bell-Evans model. It assumes first-order irreversible rate equation for the survival probability[5]. In this model, empirically determined, exponentially force-dependent unfolding rate[6–8] is used. Evans and Ritchie derived the probability density function (PDF) of the unfolding force, and they also obtained the most probable unfolding force depending on the loading rate and the two parameters: the unfolding rate and the distance between the native and the transition states. The dependence of the mean unfolding force on the loading rate for this model Eq. (1) was obtained later[9]:

$$\langle F \rangle = \frac{k_B T}{x^\ddagger} e^{\frac{k_0 k_B T}{\dot{F} x^\ddagger}} E_1\left(\frac{k_0 k_B T}{\dot{F} x^\ddagger}\right) \qquad (1)$$

where $\langle F \rangle$ is the mean unfolding force, $k_B$ is the Boltzmann constant, $T$ is the absolute temperature, $x^\ddagger$ is the distance between the native state and the transition state, $k_0$ is the unfolding rate in the absence of the external force, $\dot{F} \equiv dF/dt$ is the loading rate, and $E_1(\ )$ is the exponential integral[10,11].

Here we would like to emphasize that the determination of the mean unfolding force from the experimentally obtained unfolding force distribution is less error-prone and more straightforward compared to the determination of the most probable unfolding force. Typically, the most probable unfolding force is determined by means of fitting the Gaussian to the unfolding force distribution[12,13] and then taking the maximum of that fit as the most probable unfolding force, while neglecting the fact that generally the unfolding force distributions are asymmetric[14–18].

The biomolecular (un)folding could be viewed as diffusion in the multidimensional potential energy surface. Each configuration of the biomolecule will be represented by a point in its degree of freedom space. For each configuration, there is a corresponding potential energy, consequently, the potential energy axis could be added as an additional dimension to the degree of freedom space. Once the potential energy axis is added as described, in this newly created space of degrees of freedom and the potential energy, we will have the above-mentioned potential energy surface. In this framework, the continuous changes of the biomolecular configuration due to the thermal motion could be viewed as a Brownian motion of the particle in the degree of freedom space, in the presence of the force: the gradient of the potential energy. Under certain assumptions, this multidimensional potential energy surface could be coarse-grained into a one-dimensional free energy profile[19]. The use of Brownian dynamics (BD) simulations for mimicking dynamics of biomolecules is originates from this framework.

Kramers assumed a simple scenario with a smooth energy profile which had a well and a barrier. The particle is initially at the well (protein is folded at its native state), then due to the Brownian motion the particle crosses the barrier (protein unfolds). Kramers addressed the determination of the rate of this reaction[20]. He managed to solve the problem for large and small limiting cases of the viscosity (generally, for biomolecular systems the limit of large viscosity is relevant) assuming that the barrier height is much larger than the thermal fluctuation energy ($k_\mathrm{B}T$).

Kramers' result was used for the development of the next widely used DFS model, i.e., the profile assumption model commonly referred to as Dudko-Hummer-Szabo model[15]. Here two free energy profiles have been assumed: parabolic-cusp[14] and linear-cubic. The force-dependent unfolding rate was calculated based on Kramers' result[20,21]. Assuming first-order irreversible rate equation for the survival probability, the approximate unfolding force PDF and the approximate mean unfolding force Eq. (2) (that was further improved[22] Eq. (3)) were derived for this model. These were derived depending on the loading rate and three main parameters: the unfolding rate, the distance between the native state and the transition state, and the free energy difference between the transition state and the native state. One of the important achievements of this study was that the authors managed to combine the results of two different free energy profiles into a single expression of the unfolding force PDF (and hence its moments). As a result of this, the fourth parameter that specifies the free energy profile is added. It is equal to 1/2 for harmonic-cusp free energy profile and 2/3 for liner-cubic free energy profile. When that parameter takes value 1, the phenomenological model is recovered. The profile assumption model is applicable for the loading rates small enough to allow the unfolding to occur before Kramers' high energy barrier approximation breaks down[20,21].

$$\langle F \rangle \approx \frac{\Delta G^{\ddagger}}{\nu x^{\ddagger}} \left( 1 - \left( \frac{k_B T}{\Delta G^{\ddagger}} \ln \left( \frac{k_0 k_B T e^{\frac{\Delta G^{\ddagger}}{k_B T} + \gamma}}{\dot{F} x^{\ddagger}} \right) \right)^{\nu} \right) \qquad (2)$$

$$\langle F \rangle \approx \frac{\Delta G^{\ddagger}}{\nu x^{\ddagger}} \left( 1 - \left( 1 - \frac{k_B T}{\Delta G^{\ddagger}} e^{\frac{k_0 k_B T}{\dot{F} x^{\ddagger}}} E_1 \left( \frac{k_0 k_B T}{\dot{F} x^{\ddagger}} \right) \right)^{\nu} \right) \qquad (3)$$

In Eq. (2) and Eq. (3), $\Delta G^{\ddagger}$ is the free energy difference between the transition state and the native state, $\nu$ is the free energy profile defining factor, and $\gamma \approx 0.5772$ is the Euler-Masceroni constant.

The profile assumption model was further generalized for the case of a larger number of underlying free energy profiles[23]. In this model, the fourth parameter $\mu$ (similar to $\nu$ in the previous model[15]) determines the free energy profile, and it assumes any value between 0 and 1. It gives the same results as the previous model[15] for values 1/2, 2/3, and 1. The mean unfolding force dependence on the loading rate for this model is given by Eq. (4):

$$\langle F \rangle \approx \frac{\Delta G^{\ddagger}}{\mu x^{\ddagger}} \left( 1 - \left( \frac{k_B T}{\Delta G^{\ddagger}} \ln(\Lambda) \right) \right)^{\mu} \left( 1 + \frac{\mu(3\mu - 2) \ln(\ln(\Lambda))}{\ln(\Lambda)} \right) \qquad (4)$$

where $\Lambda \equiv \dfrac{k_0 k_B T \left( \dfrac{\Delta G^{\ddagger}}{k_B T} \right)^{2-3\mu} e^{\frac{\Delta G^{\ddagger}}{k_B T} + \gamma}}{x^{\ddagger} \dot{F}}$ .

The model that describes relatively well the unfolding force distributions for the low and high loading rates is commonly referred to as Bullerjan-Sturm-Kroy model. Here we will refer to it as a rapid model[18]. Authors managed to calculate the PDF for the unfolding forces and the mean unfolding force dependence on the loading rate Eq. (5). In this model, the mean unfolding force is simply the sum of the mean unfolding force predicted by the phenomenological model and the mean unfolding force for the limit of high loading rates when the unfolding is ballistic and follows the $\sqrt{\dot{F}}$ law[14].

$$\langle F \rangle \approx \frac{k_B T}{x^{\ddagger}} e^{\frac{k_0 k_B T}{\dot{F} x^{\ddagger}}} E_1\left(\frac{k_0 k_B T}{\dot{F} x^{\ddagger}}\right) + \sqrt{2\zeta \dot{F} x^{\ddagger}} \qquad (5)$$

In Eq. (5), $\zeta$ quantifies the friction. This friction could be the combination of the solvent viscosity, internal friction due to interatomic interactions, and other factors that are damping the fluctuations of the molecule[24]. This friction is related to the diffusion coefficient as $\zeta = k_B T / D$. This model assumes harmonic-cusp free energy profile. Using Kramers' high friction approximation, the intrinsic unfolding rate is expressed through the other parameters of the system in the following equation: $k_0 = \frac{2\Delta G^{\ddagger}}{\zeta x^{\ddagger 2}} \sqrt{\frac{\Delta G^{\ddagger}}{\pi k_B T}} e^{-\frac{\Delta G^{\ddagger}}{k_B T}}$ [18].

For sufficiently low loading rates, in the close to equilibrium regime, before the final unfolding, the system might hop back and forth between various states[25–30]. All the above-discussed models neglect the possibility of refolding. If hopping is observed they only use the first unfolding force. However, there is some information on the equilibrium properties of the system stored in the data which is obtained from close to equilibrium measurements. This can give a more complete picture of the system under study.

The model that considers also the refolding rate in the rate equation for the survival probability is frequently referred to as Friddle-Noy-De Yoreo model. We will refer to it as a reversible model[31–33]. As the phenomenological model[5], this model assumes exponentially force-dependent rates[6–8], and it considers that the force is applied by means of the Hookean spring (henceforth, spring). For this model, the dependence of the mean unfolding force on the loading rate and on three parameters was obtained Eq. (6)[31–33]. The parameters are the following: the unfolding rate, the distance between the native state and the transition state, and the free energy difference between the unfolded state and native state.

$$\langle F \rangle \approx F_{eq} + \frac{k_B T}{x^{\ddagger}} e^{\frac{k(F_{eq})k_B T}{\dot{F} x^{\ddagger}}} E_1\left(\frac{k(F_{eq})k_B T}{\dot{F} x^{\ddagger}}\right) \qquad (6)$$

In Eq. (6), $F_{eq}$ is the equilibrium force which is close to the force at which unfolding and refolding rates are having the same value (assuming that in the absence of force the unfolding rate is lower than the refolding rate). In this model, $F_{eq} = \sqrt{2\kappa \Delta G}$, where $\kappa$ is the spring constant, $\Delta G$ is the free energy difference between the unfolded state and the native state, and $k(F_{eq}) = k_0 e^{\frac{F_{eq} x^{\ddagger} - \frac{\kappa x^{\ddagger 2}}{2}}{k_B T}}$. The unfolding force PDF for this model has not been reported yet. As emphasized in[33], the unfolding force here should be calculated based on the total time (residence time or occupation time) spent in the native state's well before the final unfolding: loading rate multiplied by the residence time. In other words, the unfolding force in this model is the total force applied on the native state while the system is in that state's well.

Some of the above-discussed models, which take only the first unfolding force, have been extended for the case when the force is applied through the spring. For the phenomenological model, in the case when the force is applied through the spring, the expression of the mean unfolding force Eq. (7) could be obtained using the following rate expression $k(F) = k_0 e^{\frac{F x^{\ddagger} - \frac{\kappa x^{\ddagger 2}}{2}}{k_B T}}$.

$$\langle F \rangle = \frac{k_B T}{x^{\ddagger}} e^{\frac{k_0 k_B T e^{-\frac{\kappa x^{\ddagger 2}}{2 k_B T}}}{\dot{F} x^{\ddagger}}} E_1\left(\frac{k_0 k_B T e^{-\frac{\kappa x^{\ddagger 2}}{2 k_B T}}}{\dot{F} x^{\ddagger}}\right) \qquad (7)$$

In the case when force is applied through the spring, the dependence of the mean unfolding force on the loading rate for the profile assumption model is given in Eq. (8)[4,34]:

$$\langle F \rangle \approx \frac{\Delta G^{\ddagger} Z}{\nu x^{\ddagger}}\left(1 - \left(1 - \frac{k_B T}{\Delta G^{\ddagger} Z^3} e^{\frac{k_0 k_B T}{Z \dot{F} x^{\ddagger}} e^{\frac{\Delta G^{\ddagger}}{k_B T}(1-Z^3)}} E_1\left(\frac{k_0 k_B T}{\dot{F} x^{\ddagger}} e^{\frac{\Delta G^{\ddagger}}{k_B T}(1-Z^3)}\right)\right)^{\nu}\right) \qquad (8)$$

$$Z = 1 + \frac{\kappa}{\kappa_{G_0(x_{\min})}} \quad \text{and} \quad \kappa_{G_0(x_{\min})} \equiv \left.\frac{\partial^2 G_0(x)}{\partial x^2}\right|_{x=x_{\min}} \tag{9}$$

where $\kappa_{G_0(x_{\min})}$ is the curvature of the molecular free energy profile at the native state as defined in Eq. (9). $G_0(x) = -\Delta G^\ddagger \frac{v}{1-v}\left(\frac{x}{x^\ddagger}\right)^{\frac{1}{1-v}} + \frac{\Delta G^\ddagger}{v}\frac{x}{x^\ddagger}$ is the the free energy profile[15].

For the rapid model, when force is applied through the spring, Eq. (10) is reported in the original study[18]:

$$\langle F \rangle \approx \frac{k_B T}{Z x^\ddagger} e^{\sqrt{Z}\frac{k_{0un} k_B T e^{\frac{\Delta G^\ddagger}{k_B T}(1-Z)}}{\dot{F} x^\ddagger}} E_1\left(\sqrt{Z}\frac{k_{0un} k_B T e^{\frac{\Delta G^\ddagger}{k_B T}(1-Z)}}{\dot{F} x^\ddagger}\right) + \sqrt{2\zeta \dot{F} x^\ddagger} \tag{10}$$

This model assumes a harmonic-cusp free energy profile, hence, using Eq. (9) we obtain $Z = 1 + \frac{\kappa x^{\ddagger 2}}{2\Delta G^\ddagger}$.

The main aim of this study is to show the importance of the definition of the unfolding force and, in the context of various DFS models, to compare the predictions of the dependence of the mean unfolding force on the loading rate. Importantly, we provide a new model (unified model) based on the profile assumption[15,22] model, the reversible[32,33] model, and the rapid[14,18] model. The new unified model is applicable for the whole range of the loading rates from near-equilibrium to ballistic unfolding.

## METHODS

### Brownian dynamics simulations

Two sets of BD simulations were carried out. In the first set, the quartic free energy profile was ramped with constant loading rates ranging from 10 pN/s to $10^{12}$ pN/s to the final force, which was 15 pN for low loading rates (Eq. (11)). In the second set of simulations, the Morse free energy profile was pulled through a spring with a spring constant of 100 pN/nm with a constant velocity ranging from 1000 nm/s to $10^{14}$ nm/s to the final force, which was 150 pN for the low loading rates (Eq. (12)). The friction coefficient was chosen to be $10^{-5}$ pN s/nm[35] in both sets of simulations, and the overdamped Langevin equation (Eq. (13)) was solved numerically using Euler–Maruyama method.

$$G(x,t) = 100\left[\left(\frac{x}{5}\right)^4 - \left(\frac{x}{5}\right)^2\right] + 4x - \dot{F}tx \qquad (11)$$

$$G(x,t) = 41\left(1 - e^{-2\left(\frac{x}{0.1}-1\right)}\right)^2 - 41 + \frac{\kappa}{2}(Vt - x + 0.1)^2 \qquad (12)$$

In Eq. (11) and Eq. (12), $t$ is the time and $V$ is the speed with which spring is pulled.

$$\zeta \dot{x}(t) = -\frac{\partial G(x,t)}{\partial x} + \sqrt{2\zeta k_B T} R(t) \qquad (13)$$

In Eq. (13), $R(t)$ is a delta-correlated Gaussian white noise with 0 mean. The simulation time step was chosen to be $10^{-9}$ s for the quartic free energy profile and $10^{-12}$ s for the Morse free energy profile, securing less than 10 % of error due to finite time step[36]. For high loading rates, where the simulation time becomes quite short, the time steps were decreased even more to ensure that at least half a million steps were in a trajectory (based on random walk as a limit of Brownian motion, Donker's theorem). For each loading rate, 3000 trajectories were generated. For each trajectory, the initial value for the reaction coordinate was chosen randomly from the Boltzmann distribution for the corresponding free energy profile's native state neighborhood. The first, reversible, and last unfolding forces have been recorded for each trajectory and the mean unfolding forces were calculated from that data.

**Determination of the parameters**

To compare how well the models predict the outcome of BD simulations, was necessary to use the same parameters in the models that were used in the BD simulation. For the BD simulations with the quartic free energy profile, most of the parameters could be determined straightforwardly. The determination of the more "complicated" parameters is described here. The energy profile parameter ($\mu$) for the profile assumption (2016) model[23] was determined by fitting the free energy profile model proposed in[23] to the part of the quartic free energy profile. The unfolding rate was determined using Eq. (14), Fig. 1[21,37–39].

$$k = \frac{k_\mathrm{B} T}{\zeta} \cdot \frac{1}{\int_{x_{\min}}^{b} e^{\frac{G(y)}{k_\mathrm{B} T}} \int_{a}^{y} e^{-\frac{G(z)}{k_\mathrm{B} T}} dz\, dy} \qquad (14)$$

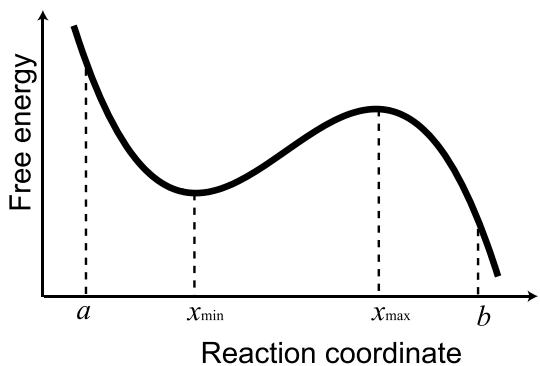

Figure. 1. The unfolding rate can be calculated using Eq. (14) where the integration limits are shown in this figure. The refolding rate can be calculated analogously.

In Eq. (14), $G(\ )$ is the free energy profile and $a$ and $b$ are reflecting and absorbing boundaries, respectively. The refolding rate was calculated using a similar equation. The equilibrium force was determined by assuming the force dependence of the free energy profile in Eq. (14) and numerically solving the equation of the unfolding and refolding rates in order to find the force at which they are equal.

In the case of the Morse free energy profile, the determination of the parameters is less trivial since this profile does not have any barrier in the absence of the external force. In order to find the distance between the native and transition states for this model, we choose as a barrier position 0.262 nm, which is the mean of the barrier positions when it first appears at 0.135 nm and when it disappears at 0.389 nm. Since the native state is at 0.1 nm, then $x^\ddagger \approx 0.162$ nm. The unfolded state distance was chosen at 0.6 nm, which is where the second minimum becomes noticeable. Therefore, for this case the first unfolding force was taken as the product of the time at which system visited the unfolded state (0.6 nm) for the first time and the loading rate, the last unfolding force was taken as the product of the time at which the system was in the native state (0.1 nm) for the last time and the loading rate, and the reversible unfolding force was taken as the product of the loading rate and the time the system spent in the native state's well (time spent for $x < 0.262$ nm). To determine the distance between the unfolded state and the transition state $x_{\text{ref}}^{\ddagger}$ and the free energy difference between the transition state and the unfolded state $\Delta G_{\text{ref}}^{\ddagger}$, for the Morse free energy profile, we followed the approach described in[40]. To determine the unfolding and refolding rates we used Eq. (14) and Fig. 1. The equilibrium force was determined with the same method as in the case of the quartic free energy profile. The determination of the parameter values for the Morse free energy profile is challenging since some parameters are virtual. Our choice of the Morse free energy profile with a spring for this study is due to the historical reasons[32]. We could have chosen a different path for determining the parameters of the Morse free energy profile, particularly by evaluating the first and third derivatives at the inflection point as described in[15]. However, parameters determined with this approach might put profile assumption models and related models in a favorable position. Ideally $\kappa V/Z$ should be used to calculate the loading rate[4,34]. However, we have used $\kappa V$ since in our case $Z \approx 1.00305$ which means that the soft spring approximation is applicable.

**$R^2$ and $\chi^2$ Criterions**

As a goodness of prediction criterion, we used $R^2$ as defined by Eq. (15).

$$R^2 \equiv 1 - \frac{\sum_{i=1}^{n}(O_i - E_i)^2}{\sum_{i=1}^{n}(O_i - \langle O \rangle)^2} \qquad (15)$$

In Eq. (15), $O_i$ is the observed value of the i$^{\text{th}}$ data point, i.e. the value we get from the data, $E_i$ is the expected value for the i$^{\text{th}}$ data point, i.e. the value predicted by the model, and $\langle O \rangle$ is the mean of the data (see Fig. 2a). Clearly, the closer the $R^2$ value is to 1, the better the model predicts the data. A model with more parameters has an advantage when models are being fitted. In such a case the adjusted $R^2$ might be used, which is defined by Eq. (16).

$$\bar{R}^2 \equiv 1 - (1 - R^2)\frac{n-1}{n-p-1} \qquad (16)$$

In Eq. (16), $R^2$ is defined by Eq. (15), $n$ is the sample size (number of data points), and $p$ is the number of parameters in the model.

Another criterion used in this study for the goodness of the prediction is $\chi^2$ defined by Eq. (17).

$$\chi^2 \equiv \sum_{i=1}^{n} \frac{(O_i - E_i)^2}{\sigma_i^2} \qquad (17)$$

In Eq. (17), $\sigma_i$ is the standard deviation for each data point and the summation goes through all the data points from first to the last ($n^{th}$) (see Fig. 2b). It is important to note that instead of the standard deviation, the standard error or other types of errors can also be used to define $\chi^2$. However, since all our data points are obtained from the averaging of the same number of unfolding forces, we use the standard deviation. Clearly, the closer $\chi^2$ is to 0, the better the model predicts the data. Similar to adjusted $R^2$, there is also a reduced $\chi^2$ that takes into account the degrees of freedom of the model, as defined by Eq. (18).

$$\bar{\chi}^2 \equiv \frac{\chi^2}{n-p-1} \qquad (18)$$

In Eq. (18), $\chi^2$ is given by Eq. (17). In our analyses we have not used the adjusted $R^2$ or reduced $\chi^2$ since we did not fit models to the data, i.e., we did not change the parameters of the models to optimize $R^2$ or $\chi^2$. By using the values in Table 1 and Table 2 and by determining the number of the data points for each range from Fig. 5 and Fig. 6, the adjusted $R^2$ and reduced $\chi^2$ can easily be calculated for all models discussed in this work using Eq. (16) and Eq. (18).

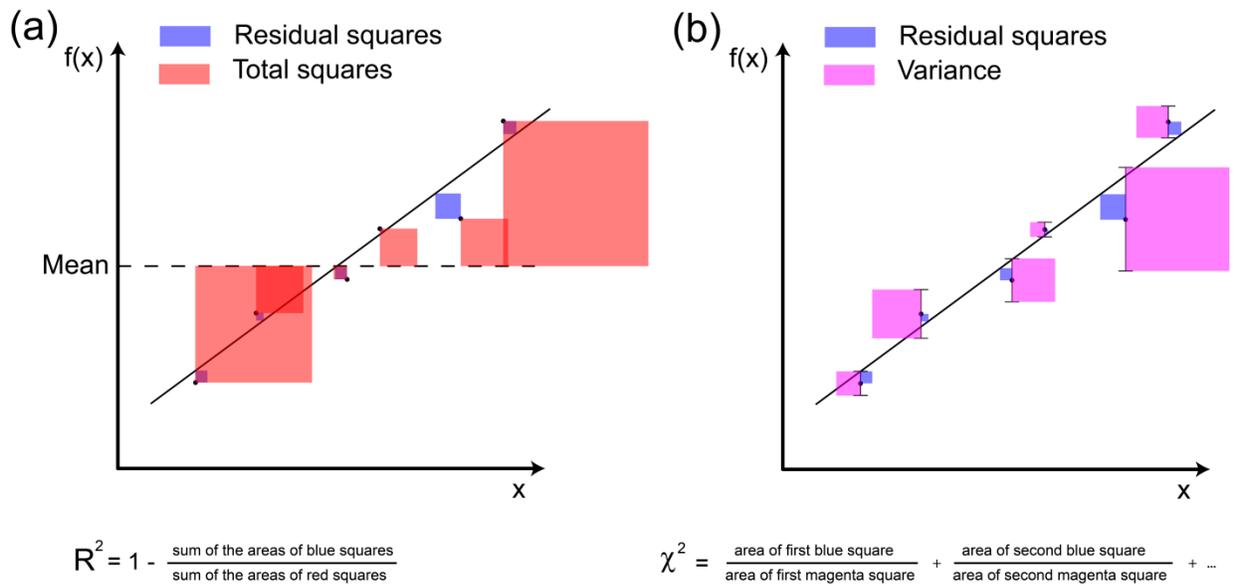

*Figure. 2. Visualization of the goodness of prediction criteria $R^2$ Eq. (15) (a) and $\chi^2$ Eq. (17) (b) used in this study. Error bars in (b) show the standard deviations. The closer the $R^2$ is to 1, the better the prediction. The closer the $\chi^2$ is to 0, the better the prediction.*

## RESULTS

### Definition of the unfolding force

The determination of the unfolding force for the case of a single transition (Fig. 3) is straightforward. However, it is unclear what force should be taken as an unfolding force if the system hops back and forth between various states before the final unfolding (Fig. 4). For such situations we define three unfolding forces. The first unfolding force is the product of the loading rate and the time when the system visits the unfolded state for the first time. The last unfolding force is the product of the loading rate and the time when the system was in the native state for the last time. The reversible unfolding force is the product of the loading rate and the time that the system spent in the native state's well (see yellow stripes in Fig. 4).

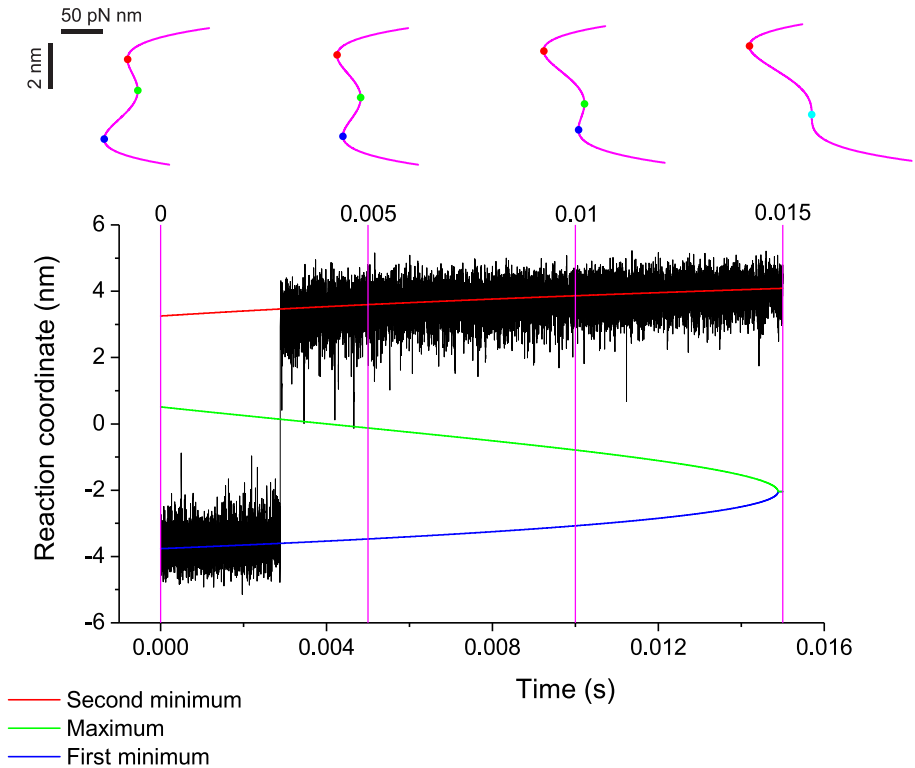

*Figure. 3. An example of a trajectory from BD simulation with a single transition. The quartic free energy profile, which is under constantly increasing external force (with 1000 pN/s loading rate), was used. There is a single transition here, and, hence, no ambiguity in the determination of the unfolding force. The graphs of the free energy profile on the top are given at different times (0.005 seconds apart) showing that the first minimum (blue dot) and the maximum (green dot) are getting closer with time and eventually merge (cyan dot). The blue, green, and red lines are showing the positions of the first minimum, maximum, and the second minimum respectively as functions of time.*

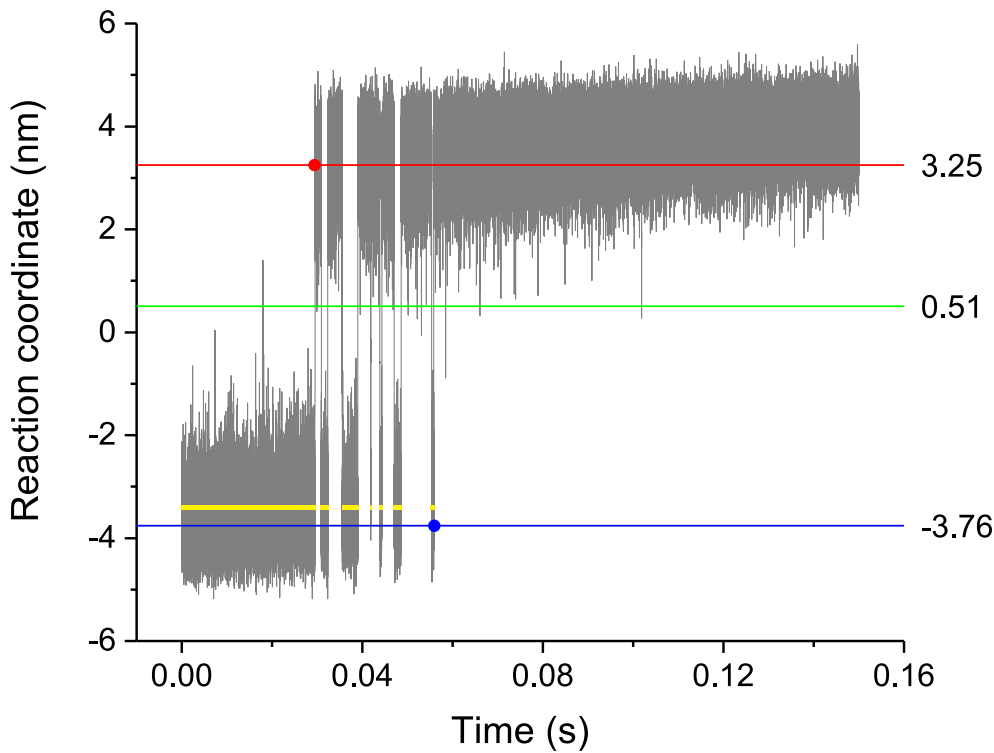

*Figure. 4. An example of a trajectory from BD simulation with multiple transitions between the native and unfolded states. The quartic free energy profile that is under constantly increasing external force (with 100 pN/s loading rate) was used. We define three unfolding forces. The first unfolding force is the product of the loading rate and the time when system visits the unfolded state (x = 3.25 nm) for the first time (red dot). The last unfolding force is the product of the loading rate and the time when the system was in the native state (x = -3.76 nm) for the last time (blue dot). The reversible unfolding force is the product of the loading rate and the time that the system spent in the native state's well (the time spent below the green line x < 0.51 nm, yellow strip). The blue, green, and red lines are showing the positions of the native, transition, and unfolded states respectively.*

As can be seen from Fig. 3, the positions of the extrema of the free energy profile change with time if the external, time-dependent force is applied. However, the native, transition, and unfolded states are specified as particular values of the reaction coordinate and are independent of the externally applied force. In the above-stated example the native, transition, and unfolded states coincide in the absence of the external force with the first minimum, maximum, and the second minimum of the free energy profile. The state is specified by a particular value (or range of values) of the reaction coordinate, and it does not depend on the external force. What depends on the external force is the likelihood of finding the system in a particular state.

In order to see how the means of these three unfolding forces depend on the loading rate, we carry out BD simulations using quartic free energy profile under the external force that increases with a constant loading rate. We generate 3000 trajectories for each loading rate, for the loading rates that span 12 orders of magnitude (see Fig. 5). For further details, see the methods section.

The surprising observation is that the mean last unfolding force shows non-monotonic dependence on the loading rate. At sufficiently low loading rates, the mean last unfolding force starts to increase with the decrease of the loading rate, a phenomenon that was not observed previously. This means that the time when the system was in the native state for the last time increases faster than the loading rate decreases, as a result the product of two increases. The behavior of the mean last unfolding force at even lower loading rates will be discussed at the end of this section (see Fig. 8). The mean reversible unfolding force at the low loading rate regime with the decrease of the loading rate tends to the constant value of $F_{eq}$. This means that the mean time spent in the native state's well increases to the same extent as the loading rate decreases. The decrease in the mean first unfolding force at the low loading rate regime is evident. As a reminder, the first unfolding force is the time when the first unfolding occurred multiplied by the loading rate. In the absence of the external force, there is a constant mean time (mean first passage time) when the system visits the unfolded state for the first time. In the presence of the slowly changing external force, the time when the first unfolding occurred tends to the mean first passage time. Hence, one of the two terms in the product tends to the constant while the other one decreases. This results in the decrease of the first unfolding force with the decreasing loading rate at the near-equilibrium regime.

As can be seen in Fig. 5, three differently defined mean unfolding forces converge with the increase of the loading rate. This is the result of moving away from the equilibrium and entering the regime where there are mostly single transitions in trajectories. In the case of a single transition, the difference between three unfolding forces is negligible as long as the time spent during the transition (transition path time) is negligible compared to the time spent before the transition (residence time of the native state). This is true for low loading rates for which also the hopping is common. This is not true for very high loading rates. For these rates only the mean reversible unfolding force is shown as the vast majority of trajectories at high loading rates have a single transition. For the single transition the determination of the unfolding force is straightforward: it is the time spent in the native state's well multiplied by the loading rate.

The next point that can be noticed from Fig. 5b is the kink around $5 \cdot 10^6$ pN/s of the loading rate value. After the critical value of the loading rate, the data points lie on a line with a higher slope. This is the indication of the change in the regime of unfolding. At such high loading rates, the unfolding mostly occurs after the disappearance of the barrier.

Hereby, we have specified three regimes: the low loading rate regime (close to equilibrium), the intermediate loading rate regime (non-equilibrium), and the high loading rate regime (non-equilibrium, ballistic unfolding). The smoothness of the kinks between these regimes (Fig. 5b) is due to the fact that the mean unfolding force for the given loading rate contains data from many trajectories. These trajectories can show multiple transitions between states, or can have a single transition to the unfolded state before or after the disappearance of the barrier.

**A unified model for the dependence of the mean unfolding force on the loading rate**

To accurately predict the behavior of the dependence of the mean unfolding force on the loading rate in all the three regimes, we developed a unified model. We used the force-dependent unfolding rate expression proposed in[15], and, following the method proposed in[22], we calculated the mean first unfolding force for the low and intermediate loading rates. Next, we extended that result for the case of high loading rates, when the unfolding is happening after the disappearance of the barrier[14,18]. This results in Eq. (19), the derivations of which can be found in Appendix A.

$$\langle F_{\text{First}} \rangle \approx \frac{\Delta G^{\ddagger}}{v x^{\ddagger}} \left( 1 - \left| 1 - \frac{k_{\text{B}} T}{\Delta G^{\ddagger}} e^{\frac{k_0 k_{\text{B}} T}{x^{\ddagger} \dot{F}}} E_1 \left( \frac{k_0 k_{\text{B}} T}{x^{\ddagger} \dot{F}} \right) \right|^{\nu} \right) \cdot \theta\left( \dot{F}_{\text{cFirst}} - \dot{F} \right) + \\ \left( \frac{\Delta G^{\ddagger}}{v x^{\ddagger}} - \sqrt{2 \dot{F}_{\text{cFirst}} x^{\ddagger} \zeta} + \sqrt{2 \dot{F} x^{\ddagger} \zeta} \right) \cdot \theta\left( \dot{F} - \dot{F}_{\text{cFirst}} \right) \qquad (19)$$

In Eq. (19), $\dot{F}_{\text{cFirst}} = \dfrac{k_0 k_B T e^{\frac{\Delta G^{\ddagger}}{k_B T}+\gamma}}{x^{\ddagger}}\left(1 - e^{-\frac{\Delta G^{\ddagger}}{k_B T}}\right)$, $k_0$ is not an independent parameter, and it can be calculated using the Kramers' approximation if the free energy profile is known. For example, for the linear-cubic profile ($v = 2/3$), we will have $k_0 = \dfrac{3\Delta G^{\ddagger}}{\pi \zeta x^{\ddagger 2}} e^{-\frac{\Delta G^{\ddagger}}{k_B T}}$ in contrast to Eq. (5), where the harmonic-cusp profile is assumed. Generally using the linear-cubic free energy profile is more reasonable as discussed in[15,34], and we will assume that profile in the future, i.e. $v = 2/3$, if not specified. Here $\theta(\ )$ is the Heaviside step function: $\theta(x) = \begin{cases} 1, & x > 0 \\ 0, & x \leq 0 \end{cases}$. Hereby, the dependence of the mean first unfolding force on the loading rate is determined by 3 independent parameters: $x^{\ddagger}$, $\Delta G^{\ddagger}$, and $\zeta$.

Following the same procedure, except this time calculating the mean differently[32], we obtained the mean reversible unfolding force: Eq. (20) (see Appendix A for details).

$$\langle F_{\text{Rev}} \rangle \approx \dfrac{\Delta G^{\ddagger}}{v x^{\ddagger}}\left(1 - \left|\left(1 - \dfrac{v F_{\text{eq}} x^{\ddagger}}{\Delta G^{\ddagger}}\right)^{\frac{1}{v}} - \dfrac{k_B T}{\Delta G^{\ddagger}} e^{\frac{k_0 k_B T}{x^{\ddagger} \dot{F}} e^{\frac{\Delta G^{\ddagger}}{k_B T}\left[1-\left(1-\frac{v F_{\text{eq}} x^{\ddagger}}{\Delta G^{\ddagger}}\right)^{\frac{1}{v}}\right]}} E_1\left[\dfrac{k_0 k_B T}{x^{\ddagger} \dot{F}} e^{\frac{\Delta G^{\ddagger}}{k_B T}\left[1-\left(1-\frac{v F_{\text{eq}} x^{\ddagger}}{\Delta G^{\ddagger}}\right)^{\frac{1}{v}}\right]}\right]\right|^{v}\right) \quad (20)$$

$$\theta\left(\dot{F}_{\text{cRev}} - \dot{F}\right) + \left(\dfrac{\Delta G^{\ddagger}}{v x^{\ddagger}} - \sqrt{2 \dot{F}_{\text{cRev}} x^{\ddagger} \zeta} + \sqrt{2 \dot{F} x^{\ddagger} \zeta}\right) \cdot \theta\left(\dot{F} - \dot{F}_{\text{cRev}}\right)$$

In Eq. (20), $\dot{F}_{\text{cRev}} = \dfrac{k_0 k_B T e^{\frac{\Delta G^{\ddagger}}{k_B T}+\gamma}}{x^{\ddagger}}\left(1 - e^{-\frac{\Delta G^{\ddagger}}{k_B T}\left(1-\frac{v F_{\text{eq}} x^{\ddagger}}{\Delta G^{\ddagger}}\right)^{\frac{1}{v}}}\right)$. Hereby, the dependence of the mean reversible unfolding force on the loading rate is determined by 4 independent parameters: $x^{\ddagger}$, $\Delta G^{\ddagger}$, $F_{\text{eq}}$, and $\zeta$.

Notice that the critical loading rate depends on the definition of the unfolding force. Generally, the mean reversible unfolding force riches the given force at the lower loading rate than the mean first unfolding force since by definition for the same trajectory the first unfolding force is not higher than the reversible unfolding force. With the increase of the loading rate, the number of trajectories with hopping decreases, and, due to this, the first and reversible mean unfolding forces are getting closer, but eventually the mean reversible unfolding force riches the critical force at a slightly smaller loading rate than the mean first unfolding force.

The mean last unfolding force can be determined based on the symmetry arguments. In Fig. 4 we observe that $\langle F_{Last} \rangle = \langle F_{Rev} \rangle + \langle F_x \rangle$ where $\langle F_x \rangle$ is the mean force due to the time spent in the unfolded state's well (above 0.51 nm), before the last unfolding. $\langle F_x \rangle$ equals the time spent in the unfolded state's well before the last unfolding multiplied by the loading rate. Interestingly, at close to equilibrium regime, it is possible to calculate $\langle F_x \rangle$ from the relaxation protocol, namely, $\langle F_x \rangle \approx \langle F_{FirstRef} \rangle - \langle F_{RevRef} \rangle$ where $\langle F_{FirstRef} \rangle$ is the man first refolding force, and $\langle F_{RevRef} \rangle$ is the mean reversible refolding force. Hence, the mean last unfolding force can be calculated through Eq. (21):

$$\langle F_{Last} \rangle \approx \langle F_{Rev} \rangle + \langle F_{FirstRef} \rangle - \langle F_{RevRef} \rangle \qquad (21)$$

At the non-equilibrium regime, the symmetry will break, but the Eq. (21) will still hold since in the case of a single transition the last two terms will cancel each other. $\langle F_{FirstRef} \rangle$ and $\langle F_{RevRef} \rangle$ can be calculated following the same path as for $\langle F_{First} \rangle$ and $\langle F_{Rev} \rangle$ only using the force dependent refolding rate as in[41] and different integration limits (see Appendix A for details). This results in Eq. (22) and Eq. (23):

$$\langle F_{FirstRef} \rangle \approx \frac{\Delta G^{\ddagger}_{ref}}{\nu x^{\ddagger}_{ref}} \left( \left| \left(1 + \frac{\nu F_{fin} x^{\ddagger}_{ref}}{\Delta G^{\ddagger}_{ref}}\right)^{\frac{1}{\nu}} - \frac{k_B T}{\Delta G^{\ddagger}_{ref}} e^{\frac{\Delta G^{\ddagger}_{ref}}{k_B T}\left[1-\left(1+\frac{\nu F_{fin} x^{\ddagger}_{ref}}{\Delta G^{\ddagger}_{ref}}\right)^{\frac{1}{\nu}}\right]} E_1\left(\frac{k_{0ref} k_B T}{x^{\ddagger}_{ref} \dot{F}} e^{\frac{\Delta G^{\ddagger}_{ref}}{k_B T}\left[1-\left(1+\frac{\nu F_{fin} x^{\ddagger}_{ref}}{\Delta G^{\ddagger}_{ref}}\right)^{\frac{1}{\nu}}\right]}\right) - 1 \right| \right) \cdot$$
$$\theta\left(\dot{F}_{cFirstRef} - \dot{F}\right) + \left(-\frac{\Delta G^{\ddagger}_{ref}}{\nu x^{\ddagger}_{ref}} + \sqrt{2 \dot{F}_{cFirstRef} x^{\ddagger}_{ref} \zeta} - \sqrt{2 \dot{F} x^{\ddagger}_{ref} \zeta}\right) \cdot \theta\left(\dot{F} - \dot{F}_{cFirstRef}\right)$$

$$(22)$$

In Eq. (22), $\dot{F}_{\text{cFirstRef}} = \dfrac{k_{0\text{ref}} k_B T e^{\frac{\Delta G^{\ddagger}_{\text{ref}}}{k_B T} + \gamma}}{x^{\ddagger}_{\text{ref}}} \left(1 - e^{-\frac{\Delta G^{\ddagger}_{\text{ref}}}{k_B T}\left(1 + \frac{\nu F_{\text{fin}} x^{\ddagger}_{\text{ref}}}{\Delta G^{\ddagger}_{\text{ref}}}\right)^{\frac{1}{\nu}}}\right)$, $k_{0\text{ref}} = \dfrac{3 \Delta G^{\ddagger}_{\text{ref}}}{\pi \zeta {x^{\ddagger}_{\text{ref}}}^2} e^{-\frac{\Delta G^{\ddagger}_{\text{ref}}}{k_B T}}$, $x^{\ddagger}_{\text{ref}}$ is the distance between the unfolded state and the transition state, $k_{0\text{ref}}$ is the refolding rate in the absence of the external force, $\Delta G^{\ddagger}_{\text{ref}}$ is the free energy difference between the transition state and the unfolded state, and $F_{\text{fin}}$ is the final force till which the system is pulled and from which the (hypothetical) refolding protocol starts. The final force is set by the experimental design and is not an intrinsic parameter describing the system.

$$\langle F_{\text{RevRef}} \rangle \approx \dfrac{\Delta G^{\ddagger}_{\text{ref}}}{\nu x^{\ddagger}_{\text{ref}}} \left( \left| \left(1 + \dfrac{\nu F_{\text{eq}} x^{\ddagger}_{\text{ref}}}{\Delta G^{\ddagger}_{\text{ref}}}\right)^{\frac{1}{\nu}} - \dfrac{k_B T}{\Delta G^{\ddagger}_{\text{ref}}} e^{\frac{k_{0\text{ref}} k_B T}{x^{\ddagger}_{\text{ref}} \dot{F}} e^{\frac{\Delta G^{\ddagger}_{\text{ref}}}{k_B T}\left[1 - \left(1 + \frac{\nu F_{\text{eq}} x^{\ddagger}_{\text{ref}}}{\Delta G^{\ddagger}_{\text{ref}}}\right)^{\frac{1}{\nu}}\right]}} E_1\left(\dfrac{k_{0\text{ref}} k_B T}{x^{\ddagger}_{\text{ref}} \dot{F}} e^{\frac{\Delta G^{\ddagger}_{\text{ref}}}{k_B T}\left[1 - \left(1 + \frac{\nu F_{\text{eq}} x^{\ddagger}_{\text{ref}}}{\Delta G^{\ddagger}_{\text{ref}}}\right)^{\frac{1}{\nu}}\right]}\right) \right|^{\nu} - 1 \right) \cdot$$

$$\theta\left(\dot{F}_{\text{cRevRef}} - \dot{F}\right) + \left(-\dfrac{\Delta G^{\ddagger}_{\text{ref}}}{\nu x^{\ddagger}_{\text{ref}}} + \sqrt{2 \dot{F}_{\text{cRevRef}} x^{\ddagger}_{\text{ref}} \zeta} - \sqrt{2 \dot{F} x^{\ddagger}_{\text{ref}} \zeta}\right) \cdot \theta\left(\dot{F} - \dot{F}_{\text{cRevRef}}\right)$$

(23)

In Eq. (23), $\dot{F}_{\text{cRevRef}} = \dfrac{k_{0\text{ref}} k_B T e^{\frac{\Delta G^{\ddagger}_{\text{ref}}}{k_B T} + \gamma}}{x^{\ddagger}_{\text{ref}}} \left(1 - e^{-\frac{\Delta G^{\ddagger}_{\text{ref}}}{k_B T}\left(1 + \frac{\nu F_{\text{eq}} x^{\ddagger}_{\text{ref}}}{\Delta G^{\ddagger}_{\text{ref}}}\right)^{\frac{1}{\nu}}}\right)$. In this case, $F_{\text{eq}}$ is not an independent parameter since in Eq. (22) and Eq. (23) there are parameters involved which describing the unfolded state's well. The equilibrium unfolding force can be calculated using the probability of having the system folded at a given force[33]:

$F_{\text{eq}} = \int_0^{\infty} \dfrac{1}{1 + e^{\frac{F(x^{\ddagger} + x^{\ddagger}_{\text{ref}}) - (\Delta G^{\ddagger} - \Delta G^{\ddagger}_{\text{ref}})}{k_B T}}} = \dfrac{k_B T}{x^{\ddagger} + x^{\ddagger}_{\text{ref}}} \ln\left(1 + e^{\frac{\Delta G^{\ddagger} - \Delta G^{\ddagger}_{\text{ref}}}{k_B T}}\right)$. This formula of $F_{\text{eq}}$ is model independent in the sense that we do not assume force dependent unfolding and refolding rates. By inserting Eq. (20), Eq. (22), and Eq. (23) into Eq. (21), the dependence of the mean last unfolding force on the loading rate is determined by 5 independent parameters: $x^{\ddagger}$, $\Delta G^{\ddagger}$, $x^{\ddagger}_{\text{ref}}$, $\Delta G^{\ddagger}_{\text{ref}}$, and $\zeta$. Independently of applying or not applying relaxation protocol, Eq. (21) (which includes Eq. (20), Eq. (22), and Eq. (23)) is applicable for predicting the dependence of the mean last unfolding force on the loading rate.

Exponential integral and Heaviside step function can be approximated using other well-known functions, for example, $e^x E_1(x) \approx \ln\left(1 + \frac{e^{-\gamma}}{x}\right)$ and $\theta(x) \approx \frac{1}{1+e^{-qx}}$ where $q$ is a large positive number[11]. However, for maximum precision, the fitting function files provided in the supplementary material exclude these approximations. These files can be directly used for data analysis.

**Comparing various models' predictions using simulated data**

To show how well all above-described models predict the dependence of the mean unfolding force on the loading rate, the models' predictions and the data from the BD simulations are presented on the same graph (Fig. 5). The parameters in these models are the same as in the BD simulations. To further quantify how well the models predicted the BD simulated data, $R^2$ and $\chi^2$ were calculated (see Table 1). For further details, see the methods section.

We observe that the phenomenological model[5,9] fails at high and intermediate loading rates. The main disadvantage of the profile assumption models[15,22,23] is that they fail at high loading rates. The rapid model[18], which combines the result of the phenomenological model[9] with the high loading rate result[14,18], arguably provides the best prediction among the previously proposed models. The reason for this could be that the intermediate loading rate regime range is small compared to the high loading rate regime, and more data is generated in the high loading rate regime. The loading rate range for the BD simulations is chosen based on the previous experimental and molecular dynamics simulations studies[13]. The unified first model clearly gives the overall best prediction. According to unified last model, the mean last unfolding force tends to the constant at very low loading rates as it will be shown later in Fig. 8. However, conducting simulations at very low loading rates is challenging due to long simulation times and large memory required to deal with simulated data. Hence, in the case of very low loading rates, the prediction of the unified last model needs further verification.

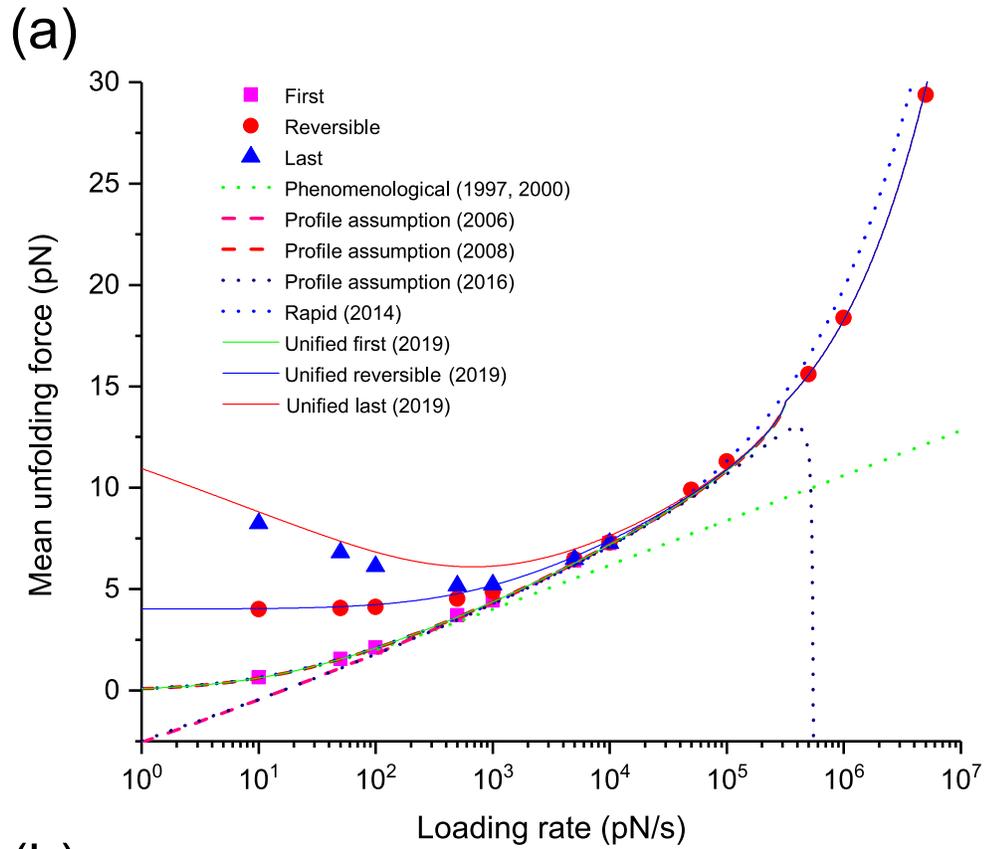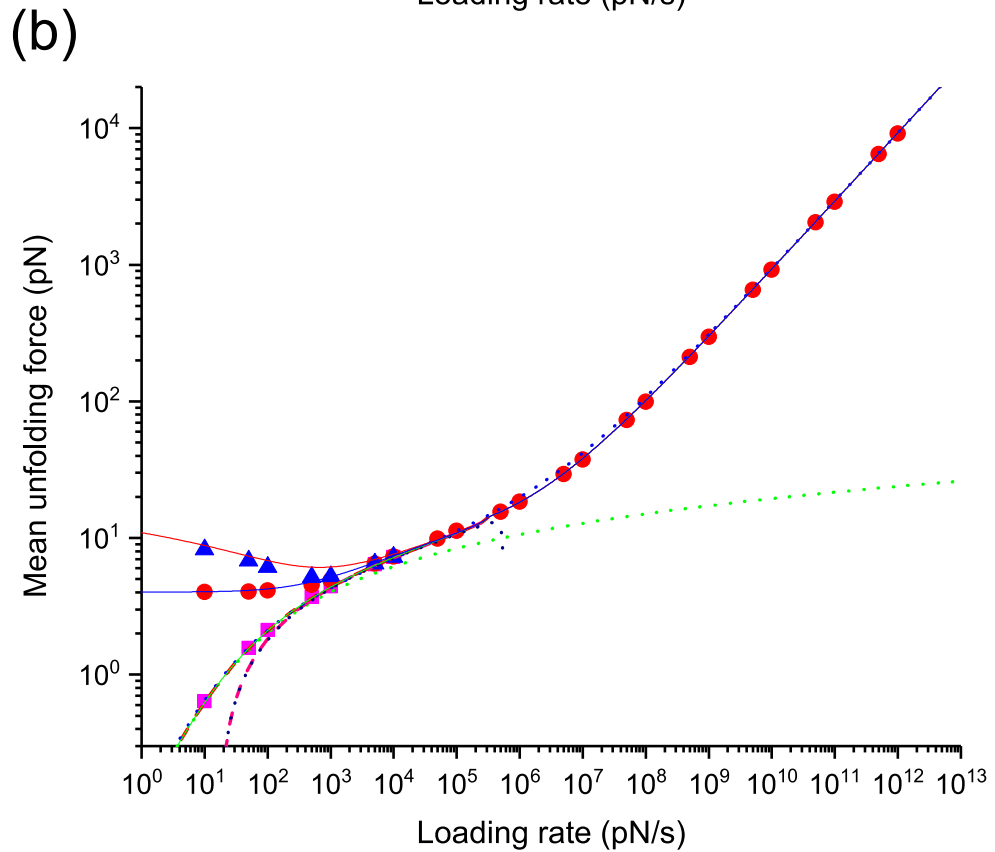

*Figure. 5. Dependence of the mean unfolding force on the loading rate, from BD simulations with quartic free energy profile, and from the predictions from the DFS models. (b) shows the log-log plot for the data in (a), but for a larger range of loading rates. Three regimes can be observed from (b). In the low loading rate regime the three differently defined mean unfolding forces diverge. This regime ends roughly where the mean last unfolding force has a minimum, the mean reversible unfolding force rises from the constant $F_{eq}$ value, and the mean first unfolding force follows a different slope, around 1000 pN/s. The intermediate loading rate regime starts where the low loading rate regime ends, and ends roughly where the change in the slope can be noticed around $10^6$ pN/s indicating that unfolding occurs after the barrier has disappeared. The models are plotted using the parameters from the BD simulation (hence these are not fits). Phenomenological (1997, 2000) corresponds to Eq. (1), profile assumption (2006) corresponds to Eq. (2), profile assumption (2008) corresponds to Eq. (3), profile assumption (2016) corresponds to Eq. (4), rapid (2014) corresponds to Eq. (5), unified first (2019) corresponds to Eq. (19), unified reversible (2019) corresponds to Eq. (20), unified last (2019) corresponds to Eq. (21) where Eq. (20), Eq. (22), and Eq. (23) were used. Table 1 shows how well these models predict the data from this figure. The parameters used in the models are $k_B T$ = 4.1 pN nm, $x^{\ddagger}$ = 4.27 nm, $k_0$ = 9.33 $s^{-1}$, $\Delta G^{\ddagger}$ = 40.62 pN nm, $\zeta$ = $10^{-5}$ pN s/nm, $F_{eq}$ = 4.02 pN, $F_{fin}$ = 15 pN, $\nu$ = 0.66667, $\mu$ = 0.5872, $x^{\ddagger}_{ref}$ = 2.74 nm, $k_{0ref}$ = 6495 $s^{-1}$, and $\Delta G^{\ddagger}_{ref}$ = 12.41 pN nm. The error bars for the standard error of the mean are smaller than the symbols.*

| Model | Low loading rates ($\dot{F}$ < 5000 pN/s) | | Intermediate loading rates (5000 pN/s ≤ $\dot{F}$ ≤ $10^6$ pN/s) | | High loading rates ($\dot{F}$ > $10^6$ pN/s) | | All loading rates | |
|---|---|---|---|---|---|---|---|---|
| | $R^2$ | $\chi^2$ | $R^2$ | $\chi^2$ | $R^2$ | $\chi^2$ | $R^2$ | $\chi^2$ |
| Phenomenological (1997) | 0.9695 | 0.1809 | 0.0275 | 12.6468 | -0.4443 | 1882.94 | -0.1878 | 1895.77 |
| Profile assumption (2006) | 0.8446 | 6.3365 | 1.0414-0.1402i | -0.432+1.4711i | -0.4441-0.0072i | 1836.41+88.0905i | -0.1876-0.0059i | 1842.31+89.5616i |
| Profile assumption (2008) | 0.9889 | 0.2245 | 1.0415-0.1402i | -0.4327+1.4711i | -0.4441-0.0072i | 1836.41+88.0905i | -0.1876-0.0059i | 1836.2+89.5616i |
| Profile assumption (2016) | 0.842 | 6.2111 | 0.883+0.028i | 1.598-0.2846i | -0.4446-0.0088i | 1824.61+107.478i | -0.188-0.0073i | 1832.42+107.193i |
| Rapid (2014) | 0.9968 | 0.0187 | 0.9727 | 0.3191 | 0.9997 | 4.145 | 0.9997 | 4.4828 |
| Unified first (2019) | 0.9987 | 0.0093 | 0.997 | 0.0693 | 0.9998 | 0.4946 | 0.9998 | 0.5732 |
| Unified reversible (2019) | 0.6972 | 0.5088 | 0.9977 | 0.0553 | 0.9998 | 0.4961 | 0.9998 | 1.0602 |
| Unified last (2019) | 0.5613 | 2.2316 | 0.9952 | 0.1886 | 0.9998 | 0.4974 | 0.9998 | 2.9176 |

*Table 1. To quantify how well the models predict the simulated data we calculate $R^2$ and $\chi^2$ for each model plotted in Fig. 5 at three regimes separately, and then all together (last column). Some of the models predict complex values for the mean unfolding force at high loading rates. For the completeness of the table, for these models, we also present $R^2$ and $\chi^2$. The perfect prediction would result in $R^2 = 1$ and $\chi^2 = 0$.*

**Unified model when the force is applied through the spring**

We further extend the unified model for the case when the system is being pulled through the spring. The dependence of the mean unfolding forces on the loading rate is derived using the same methods as when the force is applied directly. The only difference here is that different force dependent unfolding and refolding rates are used (see Appendix B for derivations)[34]. The calculations result in Eq. (24), Eq. (25), Eq. (26), and Eq. (27) for the mean first unfolding force, mean reversible unfolding force, mean first refolding force, and mean reversible refolding force respectively. The last three are used for calculating the mean last unfolding force in Eq. (21).

$$\langle F_{\text{First}}\rangle \approx \frac{\Delta G^{\ddagger}}{vx^{\ddagger}Z}\left(Z^2 - \left|Z^{\frac{2}{v}} - \frac{k_B T}{\Delta G^{\ddagger}} e^{\frac{k_0 k_B T}{x^{\ddagger} \dot{F} Z} e^{\frac{\Delta G^{\ddagger}}{k_B T}\left(1-Z^{\frac{2}{v}}\right)}} E_1\left(\frac{k_0 k_B T}{x^{\ddagger} \dot{F} Z} e^{\frac{\Delta G^{\ddagger}}{k_B T}\left(1-Z^{\frac{2}{v}}\right)}\right)\right|^v\right) \cdot \theta\left(\dot{F}_{\text{cFirst}} - \dot{F}\right) + \quad (24)$$

$$\left(\frac{\Delta G^{\ddagger} Z}{vx^{\ddagger}} - \sqrt{2\dot{F}_{\text{cFirst}} x^{\ddagger}\zeta} + \sqrt{2\dot{F} x^{\ddagger}\zeta}\right) \cdot \theta\left(\dot{F} - \dot{F}_{\text{cFirst}}\right)$$

In Eq. (24), $\dot{F}_{\text{cFirst}} = \frac{k_0 k_B T e^{\frac{\Delta G^{\ddagger}}{k_B T}+\gamma}}{x^{\ddagger} Z}\left(1 - e^{-\frac{\Delta G^{\ddagger} Z^{\frac{2}{v}}}{k_B T}}\right)$, and $Z$ is given by Eq. (9). For the linear-cubic free energy profile $Z = 1 + \frac{\kappa x^{\ddagger 2}}{6\Delta G^{\ddagger}}$.

$$\langle F_{\text{Rev}}\rangle \approx \frac{\Delta G^{\ddagger}}{vx^{\ddagger}Z}\left(Z^2 - \left|\left(Z^2 - \frac{vF_{\text{eq}} x^{\ddagger} Z}{\Delta G^{\ddagger}}\right)^{\frac{1}{v}} - \frac{k_B T}{\Delta G^{\ddagger}} e^{\frac{k_0 k_B T}{x^{\ddagger} \dot{F} Z} e^{\frac{\Delta G^{\ddagger}}{k_B T}\left(1-\left(Z^2 - \frac{vF_{\text{eq}} x^{\ddagger} Z}{\Delta G^{\ddagger}}\right)^{\frac{1}{v}}\right)}} E_1\left(\frac{k_0 k_B T}{x^{\ddagger} \dot{F} Z} e^{\frac{\Delta G^{\ddagger}}{k_B T}\left(1-\left(Z^2 - \frac{vF_{\text{eq}} x^{\ddagger} Z}{\Delta G^{\ddagger}}\right)^{\frac{1}{v}}\right)}\right)\right|^v\right) \cdot$$

$$\theta\left(\dot{F}_{\text{cRev}} - \dot{F}\right) + \left(\frac{\Delta G^{\ddagger} Z}{vx^{\ddagger}} - \sqrt{2\dot{F}_{\text{cRev}} x^{\ddagger}\zeta} + \sqrt{2\dot{F} x^{\ddagger}\zeta}\right) \cdot \theta\left(\dot{F} - \dot{F}_{\text{cRev}}\right)$$

(25)

In Eq. (25), $\dot{F}_{\text{cRev}} = \frac{k_0 k_B T e^{\frac{\Delta G^{\ddagger}}{k_B T}+\gamma}}{x^{\ddagger} Z}\left(1 - e^{-\frac{\Delta G^{\ddagger}}{k_B T}\left(Z^2 - \frac{vF_{\text{eq}} x^{\ddagger} Z}{\Delta G^{\ddagger}}\right)^{\frac{1}{v}}}\right)$.

$\langle F_{\text{Last}}\rangle$ could be calculated using Eq. (21) and Eq. (25), Eq. (26), and Eq. (27) for $\langle F_{\text{Rev}}\rangle$, $\langle F_{\text{FirstRef}}\rangle$ and $\langle F_{\text{RevRef}}\rangle$ respectively.

$$\langle F_{\text{FirstRef}} \rangle \approx \frac{\Delta G^{\ddagger}_{\text{ref}}}{v x^{\ddagger}_{\text{ref}} Z_{\text{ref}}} \left( \left[ \left( Z^2_{\text{ref}} + \frac{vF_{\text{fin}} x^{\ddagger}_{\text{ref}} Z_{\text{ref}}}{\Delta G^{\ddagger}_{\text{ref}}} \right)^{\frac{1}{v}} - \frac{k_B T}{\Delta G^{\ddagger}_{\text{ref}}} e^{\frac{k_{0\text{ref}} k_B T}{x^{\ddagger}_{\text{ref}} \dot{F} Z_{\text{ref}}} e^{\frac{\Delta G^{\ddagger}_{\text{ref}}}{k_B T}\left(1-\left(Z^2_{\text{ref}} + \frac{vF_{\text{fin}} x^{\ddagger}_{\text{ref}} Z_{\text{ref}}}{\Delta G^{\ddagger}_{\text{ref}}}\right)^{\frac{1}{v}}\right)}} E_1\left( \frac{k_{0\text{ref}} k_B T}{x^{\ddagger}_{\text{ref}} \dot{F} Z_{\text{ref}}} e^{\frac{\Delta G^{\ddagger}_{\text{ref}}}{k_B T}\left(1-\left(Z^2_{\text{ref}} + \frac{vF_{\text{fin}} x^{\ddagger}_{\text{ref}} Z_{\text{ref}}}{\Delta G^{\ddagger}_{\text{ref}}}\right)^{\frac{1}{v}}\right)} \right) \right]^v - Z^2_{\text{ref}} \right).$$

$$\theta\left(\dot{F}_{\text{cFirst}} - \dot{F}\right) + \left( -\frac{\Delta G^{\ddagger}_{\text{ref}} Z_{\text{ref}}}{v x^{\ddagger}_{\text{ref}}} + \sqrt{2 \dot{F}_{\text{cFirst}} x^{\ddagger}_{\text{ref}} \zeta} - \sqrt{2 \dot{F} x^{\ddagger}_{\text{ref}} \zeta} \right) \cdot \theta\left(\dot{F} - \dot{F}_{\text{cFirst}}\right)$$

(26)

In Eq. (26), $\dot{F}_{\text{cFirst}} = \frac{k_{0\text{ref}} k_B T e^{\frac{\Delta G^{\ddagger}_{\text{ref}}}{k_B T}+\gamma}}{x^{\ddagger}_{\text{ref}} Z_{\text{ref}}} \left( 1 - e^{-\frac{\Delta G^{\ddagger}_{\text{ref}}}{k_B T}\left(Z^2_{\text{ref}} + \frac{vF_{\text{fin}} x^{\ddagger}_{\text{ref}} Z_{\text{ref}}}{\Delta G^{\ddagger}_{\text{ref}}}\right)^{\frac{1}{v}}} \right)$, and $Z_{\text{ref}} = 1 + \frac{\kappa}{\kappa_{G_{0\text{ref}}}(x_{\min})}$ where

$\kappa_{G_{0\text{ref}}}(x_{\min}) \equiv \left. \frac{\partial^2 G_{0\text{ref}}(x)}{\partial x^2} \right|_{x=x_{\min}}$ is the curvature of the free energy profile in the unfolded state, and

$G_{0\text{ref}}(x) = \Delta G^{\ddagger}_{\text{ref}} \frac{v}{1-v} \left(\frac{x}{x^{\ddagger}_{\text{ref}}}\right)^{\frac{1}{1-v}} - \frac{\Delta G^{\ddagger}_{\text{ref}}}{v} \frac{x}{x^{\ddagger}_{\text{ref}}}$ is the model for the free energy profile. For the linear-cubic free energy profile $Z_{\text{ref}} = 1 - \frac{\kappa x^{\ddagger\,2}_{\text{ref}}}{6\Delta G^{\ddagger}_{\text{ref}}}$.

$$\langle F_{\text{RevRef}} \rangle \approx \frac{\Delta G^{\ddagger}_{\text{ref}}}{v x^{\ddagger}_{\text{ref}} Z_{\text{ref}}} \left( \left[ \left( Z^2_{\text{ref}} + \frac{vF_{\text{eq}} x^{\ddagger}_{\text{ref}} Z_{\text{ref}}}{\Delta G^{\ddagger}_{\text{ref}}} \right)^{\frac{1}{v}} - \frac{k_B T}{\Delta G^{\ddagger}_{\text{ref}}} e^{\frac{k_{0\text{ref}} k_B T}{x^{\ddagger}_{\text{ref}} \dot{F} Z_{\text{ref}}} e^{\frac{\Delta G^{\ddagger}_{\text{ref}}}{k_B T}\left(1-\left(Z^2_{\text{ref}} + \frac{vF_{\text{eq}} x^{\ddagger}_{\text{ref}} Z_{\text{ref}}}{\Delta G^{\ddagger}_{\text{ref}}}\right)^{\frac{1}{v}}\right)}} E_1\left( \frac{k_{0\text{ref}} k_B T}{x^{\ddagger}_{\text{ref}} \dot{F} Z_{\text{ref}}} e^{\frac{\Delta G^{\ddagger}_{\text{ref}}}{k_B T}\left(1-\left(Z^2_{\text{ref}} + \frac{vF_{\text{eq}} x^{\ddagger}_{\text{ref}} Z_{\text{ref}}}{\Delta G^{\ddagger}_{\text{ref}}}\right)^{\frac{1}{v}}\right)} \right) \right]^v - Z^2_{\text{ref}} \right).$$

$$\theta\left(\dot{F}_{\text{cRevRef}} - \dot{F}\right) + \left( -\frac{\Delta G^{\ddagger}_{\text{ref}} Z_{\text{ref}}}{v x^{\ddagger}_{\text{ref}}} + \sqrt{2 \dot{F}_{\text{cRevRef}} x^{\ddagger}_{\text{ref}} \zeta} - \sqrt{2 \dot{F} x^{\ddagger}_{\text{ref}} \zeta} \right) \cdot \theta\left(\dot{F} - \dot{F}_{\text{cRevRef}}\right)$$

(27)

In Eq. (27), $\dot{F}_{\text{cRevRef}} = \frac{k_{0\text{ref}} k_B T e^{\frac{\Delta G^{\ddagger}_{\text{ref}}}{k_B T}+\gamma}}{x^{\ddagger}_{\text{ref}} Z_{\text{ref}}} \left( 1 - e^{-\frac{\Delta G^{\ddagger}_{\text{ref}}}{k_B T}\left(Z^2_{\text{ref}} + \frac{vF_{\text{eq}} x^{\ddagger}_{\text{ref}} Z_{\text{ref}}}{\Delta G^{\ddagger}_{\text{ref}}}\right)^{\frac{1}{v}}} \right)$. As previously $F_{\text{eq}}$ (in the case of

the mean last unfolding force) is not an independent parameter and can be expressed through other parameters of the system[33]

$$F_{eq} = \int_0^\infty \frac{1}{1+e^{\frac{\frac{F^2}{2\kappa}-(\Delta G^\ddagger - \Delta G^\ddagger_{ref})}{k_BT}}} = -\sqrt{\frac{\pi\kappa k_B T}{2}} \text{Li}_{1/2}\left(-e^{\frac{\Delta G^\ddagger - \Delta G^\ddagger_{ref}}{k_BT}}\right) \text{ where } \text{Li}_{1/2}(\ ) \text{ is the polylogarithm[11].}$$

In order to see how well different models predict the dependence of the mean unfolding force on the loading rate when a pulling force is applied through a spring, we carried out BD simulations using Morse free energy profile which was pulled with constant speed through a spring. The spring constant was 100 pN/nm and the Morse free energy profile parameters were chosen as in[32]. We generated 3000 trajectories using loading rates, for the loading rates ranging through 11 orders of magnitude, as presented in Fig. 6 (see methods section for details).

To show how well above discussed models predict the dependence of the mean unfolding force on the loading rate, we plot the models' predictions and the data from the BD simulations on the same graph (Fig. 6). The parameters for plotting the models were the same as those used in the BD simulations. To quantify how well the models predict the BD simulation data, we have calculated $R^2$ and $\chi^2$ (see methods section for details), which are presented in Table 2. Unified models generally give the best predictions. It appears that the soft spring approximation works well in this case because $Z$ is close to 1, and hence we could have used models that do not take into account the correction for the spring. However, since the reversible model[32,33] was originally developed assuming pulling through the spring, we used models that take into account the effect of the spring in order to have all models on equal ground.

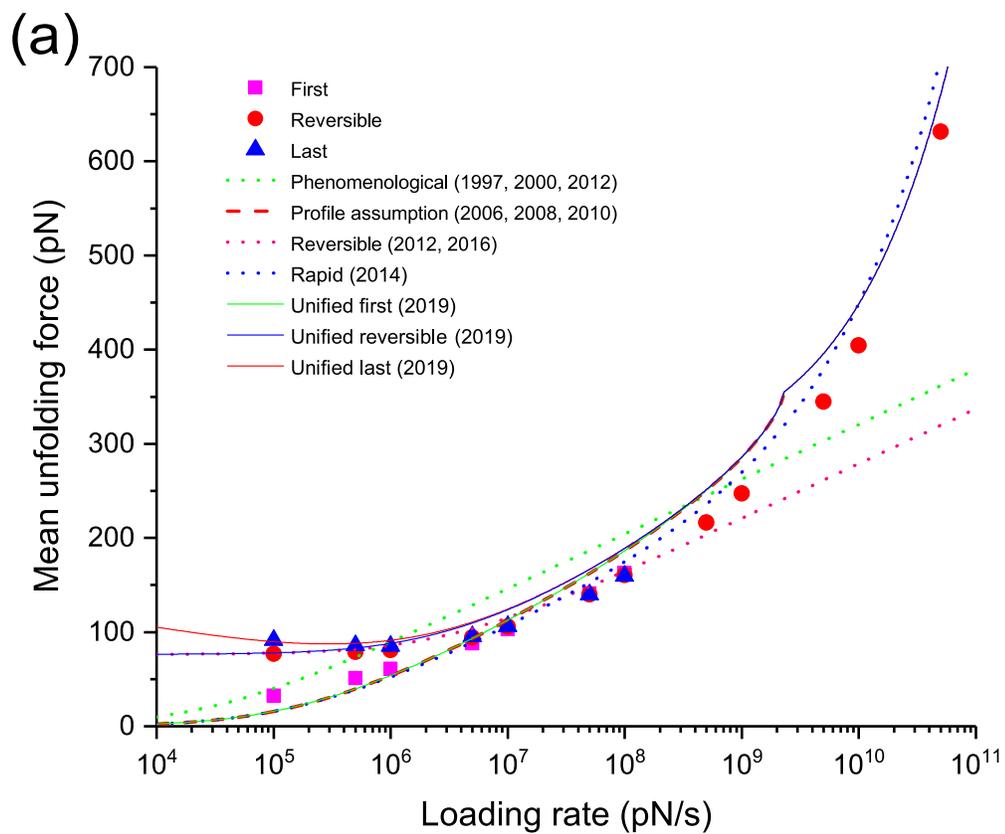

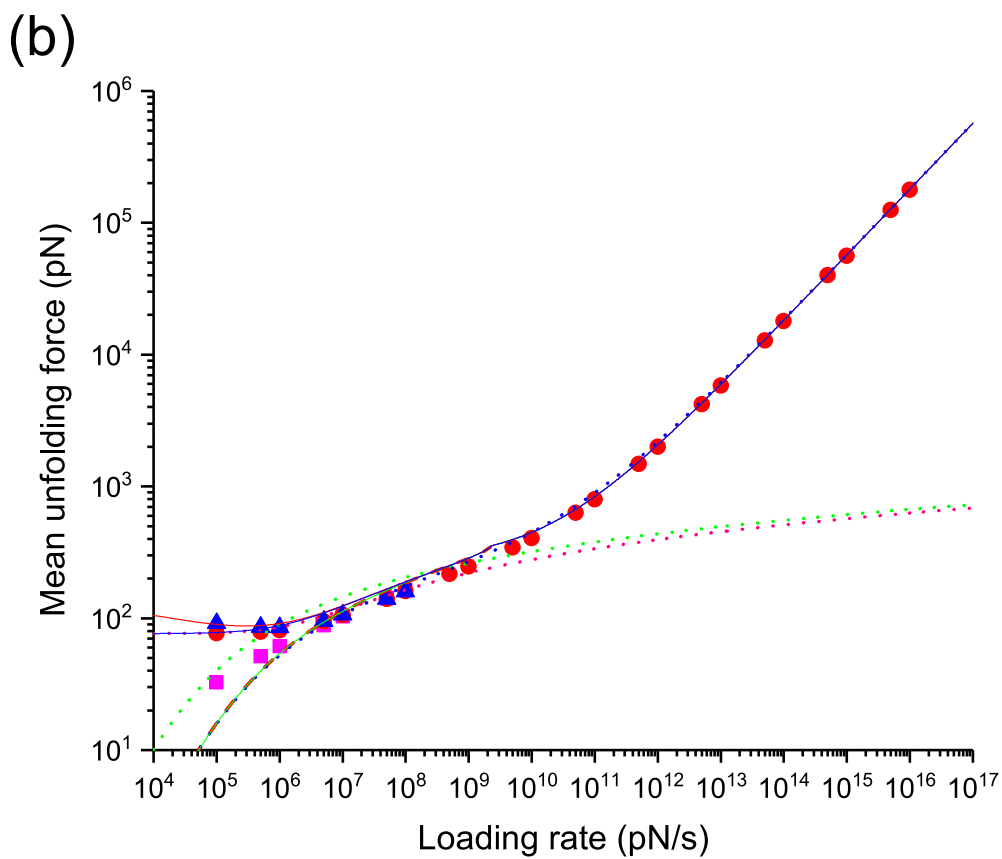

*Figure. 6. Dependence of the mean unfolding force on the loading rate, from BD simulations with Morse free energy profile, and from the predictions from the DFS models. (b) shows the log-log plot for the data in (a), but for a larger range of loading rates. Three regimes can be observed in (b). In the low loading rate regime the three differently defined mean unfolding forces diverge. This regime ends roughly where the mean last unfolding force has a minimum, the mean reversible unfolding force rises from the constant $F_{eq}$ value, and the mean first unfolding force follows a different slope, around $5 \cdot 10^6$ pN/s. The intermediate loading rate regime starts where the low loading rate regime ends and ends roughly where the change in the slope can be noticed around $10^{10}$ pN/s, indicating that unfolding occurs after the barrier has disappeared. The models are plotted using the parameters from the BD simulation (hence these are not fits). Phenomenological (1997, 2000, 2012) corresponds to Eq. (7), profile assumption (2006, 2008, 2010) corresponds to Eq. (8), reversible (2012, 2016) corresponds to Eq. (6), rapid (2014) corresponds to Eq. (10), unified first (2019) corresponds to Eq. (24), unified reversible (2019) corresponds to Eq. (25), unified last (2019) corresponds to Eq. (21) where Eq. (25), Eq. (26), and Eq. (27) were used. Table 2 shows how well models predict the data from this figure. The parameters used in the models were $k_B T = 4.1$ pN nm, $\kappa = 100$ pN/nm, $x^{\ddagger} = 0.162$ nm, $k_0 = 5002.51$ s$^{-1}$, $\Delta G^{\ddagger} = 37.85$ pN nm, $\zeta = 10^{-5}$ pN s/nm, $F_{eq} = 76.4$ pN, $F_{fin} = 150$ pN, $v = 0.66667$, $x^{\ddagger}_{ref} = 0.338$ nm, $k_{0ref} = 550713$ s$^{-1}$, and $\Delta G^{\ddagger}_{ref} = 8.4$ pN nm. The error bars for the standard error of the mean are smaller than the symbols.*

| Model | Low loading rates ($\dot{F} < 10^7$ pN/s) | | Intermediate loading rates ($10^7$ pN/s $\leq \dot{F} \leq 10^{10}$ pN/s) | | High loading rates ($\dot{F} > 10^{10}$ pN/s) | | All loading rates | |
|---|---|---|---|---|---|---|---|---|
| | $R^2$ | $\chi^2$ | $R^2$ | $\chi^2$ | $R^2$ | $\chi^2$ | $R^2$ | $\chi^2$ |
| Phenomenological (1997, 2000, 2012) | -0.8937 | 8.5561 | 0.7867 | 8.6606 | -0.4415 | 2534.49 | -0.1848 | 2550.6 |
| Profile assumption (2006, 2008, 2010) | 0.6993 | 2.5944 | 1.0765+ 0.0716i | 0.6246- 1.0892 | -0.4428- 0.0096i | 2513.46+ 112.249i | -0.1859- 0.0079i | 2517.91+ 111.189i |
| Rapid (2014) | 0.6861 | 2.726 | 0.9396 | 1.2329 | 0.9995 | 4.8118 | 0.9996 | 8.459 |
| Unified first (2019) | 0.6993 | 2.5944 | 0.8855 | 3.0696 | 0.9996 | 1.2131 | 0.9997 | 6.5727 |
| Reversible (2012, 2016) | 0.3416 | 1.2726 | 0.6724 | 4.6319 | -0.4426 | 2554.35 | -0.1858 | 2560.25 |
| Unified reversible (2019) | -0.6159 | 2.7631 | 0.8735 | 4.1891 | 0.9996 | 1.2143 | 0.9997 | 8.1665 |
| Unified last (2019) | -3.0559 | 1.5944 | 0.8723 | 4.275 | 0.9996 | 1.2144 | 0.9997 | 7.8419 |

*Table 2. To quantify how well the models predict the simulated data we calculate $R^2$ and $\chi^2$ for each model plotted in Fig. 6 at three regimes separately and then all together (last column). The profile assumption (2006, 2008, 2010) model predicts complex values for the mean unfolding force at high loading rates. For the completeness of the table we present $R^2$ and $\chi^2$ for this model as well. Perfect prediction would result in $R^2 = 1$ and $\chi^2 = 0$.*

**Distributions of differently defined unfolding forces**

The distributions of differently defined unfolding forces in the low loading rate regime are vastly different (Fig. 7). The reversible unfolding force distribution is symmetric and narrow. The last unfolding force is more widely distributed and has a right tail. Generally at intermediate loading rates, in cases of single transition, the unfolding force distribution has a left tail[14–18]. When there is hopping, the first unfolding force distribution also shows a left tail, but as the loading rate decreases the overall distribution moves closer to 0 and the left tail disappears, as can be seen in Fig. 7a. The evolution of the differently defined unfolding force distributions over the loading rate is an interesting question since the distributions are very different in the near-equilibrium regime but eventually coincide as the system moves further from equilibrium. A subject of future research could be the derivation of a differential equation describing these "dynamics".

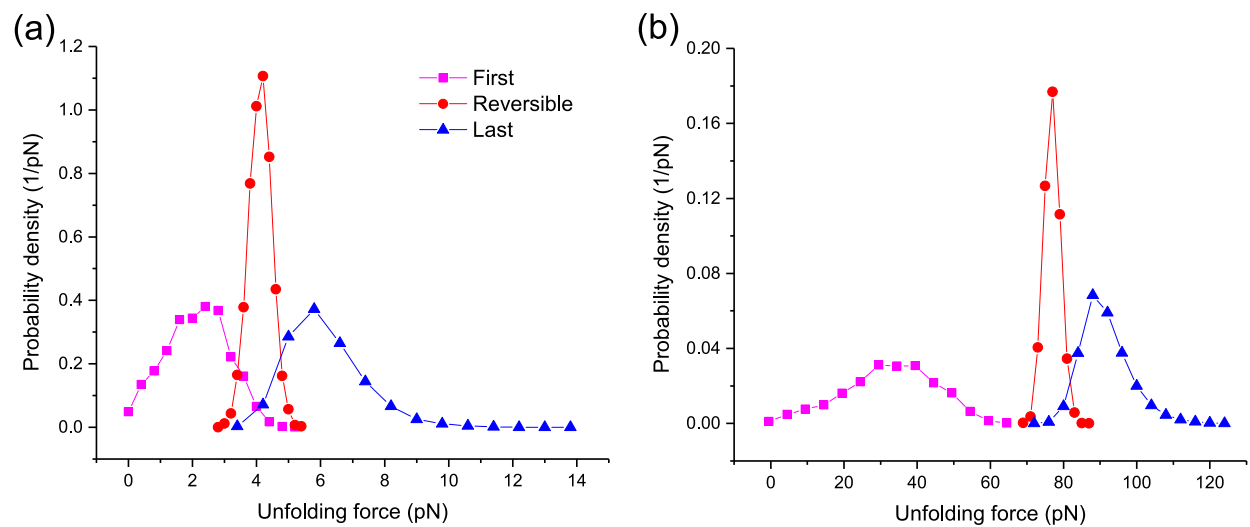

*Figure. 7. Distributions of the first, reversible, and last unfolding forces obtained from BD simulations on quartic free energy profile with a loading rate of 100 pN/s (a) and on a Morse free energy profile where the force is applied through a spring with 100 pN/nm stiffness that was pulled at a speed of 1000 nm/s (b). The last unfolding force distribution shows a right tail. The reversible unfolding force distribution is symmetric around its mean and is the most compact. The characteristic left tail of the first unfolding force distribution disappears at low loading rates.*

**Mean last unfolding force behavior at low loading rates**

As described above, in the low loading rate regime the mean last unfolding force starts to increase with the decrease of the loading rate. Performing BD simulations at lower loading rates is challenging because it requires long simulation times and large memory. However, we could see the prediction of the unified last model at lower loading rates (Fig. 8).

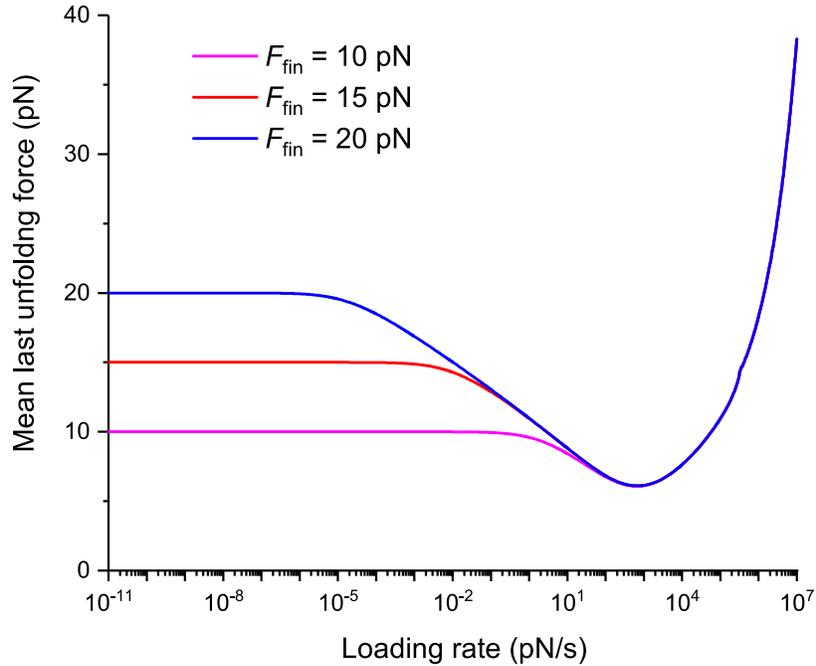

*Figure. 8. The dependence of the mean last unfolding force on the loading rate for very low loading rates from the unified last model. For very low loading rates, the mean last unfolding force approaches toward the constant value of the final force ($F_{fin}$) until which the system is driven. The final force is not an intrinsic parameter of the system but is set by the experimenter and can be higher than the critical force (force at which barrier disappears). These curves were generated using Eq. (21) with Eq. (20), eq (22), and Eq. (23), with parameter values $k_B T = 4.1$ pN nm, $x^{\ddagger} = 4.27$ nm, $k_0 = 9.33$ s$^{-1}$, $\Delta G^{\ddagger} = 40.62$ pN nm, $\zeta = 10^{-5}$ pN s/nm, $F_{eq} = 4.02$ pN, $\nu = 0.66667$, $x^{\ddagger}_{ref} = 2.74$ nm, $k_{0ref} = 6495$ s$^{-1}$, and $\Delta G^{\ddagger}_{ref} = 12.41$ pN nm.*

We first observe an increase and then a plateau of the mean last unfolding force at low loading rates. This means that the time for the last unfolding increases with decreasing loading rate, such that the product of that time and the loading rate still increases. When the loading rate goes to 0, $\langle F_{\text{FirstRef}} \rangle$ (Eq. (22)) tends toward $F_{\text{fin}}$, i.e. to the final force until which the system is pulled or the force from which the relaxation could start. At the same limit, $\langle F_{\text{Rev}} \rangle$ (Eq. (20)) and $\langle F_{\text{RevRef}} \rangle$ (Eq. (23)) both tend toward the same constant $F_{\text{eq}}$, hence $\langle F_{\text{Last}} \rangle$ (Eq. (21)) will tend toward $F_{\text{fin}}$ as $\langle F_{\text{FirstRef}} \rangle$. The plateau is an indication that at a sufficiently low loading rates hopping will continue until the end of the process, i.e. until the final force to which the system is driven. This occurs because there is always a finite probability that the system will refold back and unfold again, even if the energy of the native state has become much higher than that of unfolded state during the pulling process. Over sufficiently long time periods, such as those needed to accomplish the entire pulling process at very low loading rates, such improbable events are likely to occur.

## DISCUSSION

It is crucial to understand that different models might use different definitions of the unfolding forces. As was shown in Fig. 5 and Fig. 6, differently defined mean unfolding forces are very different in the low loading rate regime. Nevertheless, the reversible model[32,33] has been widely used for fitting the dependence of the mean unfolding force on the loading rate without specifying how the unfolding forces were determined[13,42–44]. It is possible that data was collected from the medium and high loading rate regimes where hopping is not observed. However, if there is a lack of data from all three regimes some of the parameters cannot be estimated reliably. For example, if there is no data from the low loading rate regime (where the hopping occurs) then the parameters describing the equilibrium properties of the system (such as $F_{\text{eq}}$, $x^{\ddagger} + x^{\ddagger}_{\text{ref}}$ and $\Delta G^{\ddagger} - \Delta G^{\ddagger}_{\text{ref}}$) cannot be reliably estimated. The detection of hopping is not only related to lower loading rates but also requires a higher signal to noise ratio and temporal resolution. For example in mechanical unfolding studies of bacteriorhodopsin through AFM from the freely spanning membrane which lowers the loading rates considerably, hopping was not observed[45]. However, using modified cantilevers which significantly increase the temporal resolution hopping and new intermediates in bacteriorhodopsin unfolding have been observed[30,46]. Therefore, we encourage researchers[28,30,46] who possess the data or capability of acquiring data where hopping is detected for a wide range of loading rates to test the predictions of the unified model for the mean last unfolding force.

The little cusp in the unified models at the transition from intermediate to high loading rates (see Fig. 5a and Fig. 6a) is a consequence of using the profile assumption model[15,22], which in turn uses Kramers' high barrier approximation[20]. The barrier gets lower under the externally applied force and the approximation breaks down before but close to the critical force. In the current study, the unification of the dependence of the mean unfolding force on the loading rate in the low and intermediate loading rate regimes with the dependence in the high loading rate regime was done using a mathematical trick. In order to have a full physical description, we need an analytical and analytically integrable expression for the dependence of the unfolding rate on the force. The expression must avoid the Kramers' high barrier approximation and be valid for the whole range of the forces, even above the critical force.

## CONCLUSIONS

The determination of the unfolding force from the simulated or measured data (trajectory or force-extension curve) is highly important for obtaining accurate results for the energetic and dynamic parameters of the system under study. The determination of the unfolding force should be done according to the definition of the unfolding force used in the DFS model in question, according to which dependence of the mean unfolding force on the loading rate is analyzed. This is especially important in the low loading rate regime where the three unfolding force definitions, discussed in this work, differ significantly.

The mean last unfolding force is more instructive than, for example, the mean first unfolding force. The unified last model has 5 independent parameters, and the curve describing the dependence of the mean last unfolding force on the loading rate shows more features than the curve describing the dependence of the mean first unfolding force on the loading rate.

The unified models (unified first, unified reversible, unified last) proposed in this work give better overall predictions for the BD simulated data compared to other currently available models. Many of the previously proposed models fail in the high loading rate regime. The decisive test for these models would be DFS measurements on a system for which all the energetic and dynamic parameters are known, either through the design of the system (protein) or by different direct measurements. Then, the models' predictions should be fitted to the data for the dependence of the mean unfolding force on the loading rate. The parameters obtained from those fits should be compared to the actual values of the parameters.

# APPENDIX A

**Mean reversible rnfolding force dependence on the loading rate for the unified model**

The derivation of Eq. (20) will be shown in two steps. First, the dependence of the mean unfolding force on the loading rate for the low and intermediate loading rates is calculated. Second, that result is united with the result for high loading rates.

For the first part, we follow the approach of[32] with a difference that we use the profile assumption rate expression[15] (Eq. (A1)) which is more general than the phenomenological rate[6–8] used in[32] and reduces to it when $v = 1$.

$$k(F) = k_0 \left(1 - \frac{vFx^{\ddagger}}{\Delta G^{\ddagger}}\right)^{\frac{1}{v}-1} e^{\frac{\Delta G^{\ddagger}}{k_B T}\left(1 - \left(1 - \frac{vFx^{\ddagger}}{\Delta G^{\ddagger}}\right)^{\frac{1}{v}}\right)} \quad (A1)$$

$$\int k(F) dF = \frac{k_0 k_B T}{x^{\ddagger}} e^{\frac{\Delta G^{\ddagger}}{k_B T}\left(1 - \left(1 - \frac{vFx^{\ddagger}}{\Delta G^{\ddagger}}\right)^{\frac{1}{v}}\right)} \quad (A2)$$

Using the approach of[32], by substituting Eq. (A2) into Eq. (A3) we get Eq. (A4).

$$\int_1^p \frac{dp'}{p'} = -\frac{1}{\dot{F}} \int_{F_{eq}}^F k(F') dF' \quad (A3)$$

$$\ln p = -\frac{k_0 k_B T}{x^{\ddagger} \dot{F}} \left( e^{\frac{\Delta G^{\ddagger}}{k_B T}\left(1-\left(1-\frac{\nu F x^{\ddagger}}{\Delta G^{\ddagger}}\right)^{\frac{1}{\nu}}\right)} - e^{\frac{\Delta G^{\ddagger}}{k_B T}\left(1-\left(1-\frac{\nu F_{eq} x^{\ddagger}}{\Delta G^{\ddagger}}\right)^{\frac{1}{\nu}}\right)} \right) \qquad (A4)$$

By solving Eq. (A4) for $F$ we find Eq. (A5).

$$F = \frac{\Delta G^{\ddagger}}{\nu x^{\ddagger}} \left( 1 - \left( 1 - \frac{k_B T}{\Delta G^{\ddagger}} \ln \left( e^{\frac{\Delta G^{\ddagger}}{k_B T}\left(1-\left(1-\frac{\nu F_{eq} x^{\ddagger}}{\Delta G^{\ddagger}}\right)^{\frac{1}{\nu}}\right)} - \frac{x^{\ddagger} \dot{F} \ln p}{k_0 k_B T} \right) \right)^{\nu} \right) \qquad (A5)$$

The advantage of using the rate in Eq. (A1)[15] versus the rate used in[23], where the first power is $2 - 1/\mu$ instead of $1/\nu - 1$, is that the indefinite integral in Eq. (A2) will contain an En-function[10,11] which complicates the later step of solving Eq. (A4) for $F$.

At this step, we follow the approximate averaging procedure as in[22].

$$\alpha \equiv \ln \left( e^{\frac{\Delta G^{\ddagger}}{k_B T}\left(1-\left(1-\frac{\nu F_{eq} x^{\ddagger}}{\Delta G^{\ddagger}}\right)^{\frac{1}{\nu}}\right)} - \frac{x^{\ddagger} \dot{F} \ln p}{k_0 k_B T} \right)$$

$$\langle F \rangle \approx F(\langle \alpha \rangle)$$

$$\int_0^1 \ln\left( A - \frac{\ln p}{B} \right) dp = e^{AB} E_1(AB) + \ln(A) \approx \ln\left( A + \frac{e^{-\gamma}}{B} \right) \qquad (A6)$$

From Eq. (A5) and Eq. (A6), we obtain Eq. (A7) for the mean reversible unfolding force.

$$\langle F_{Rev} \rangle \approx \frac{\Delta G^{\ddagger}}{\nu x^{\ddagger}} \left( 1 - \left( 1 - \frac{k_B T}{\Delta G^{\ddagger}} \left( e^{\frac{k_0 k_B T}{x^{\ddagger} \dot{F}} e^{\frac{\Delta G^{\ddagger}}{k_B T}\left(1-\frac{\nu F_{eq} x^{\ddagger}}{\Delta G^{\ddagger}}\right)^{\frac{1}{\nu}}}} E_1\left( \frac{k_0 k_B T}{x^{\ddagger} \dot{F}} e^{\frac{\Delta G^{\ddagger}}{k_B T}\left(1-\left(1-\frac{\nu F_{eq} x^{\ddagger}}{\Delta G^{\ddagger}}\right)^{\frac{1}{\nu}}\right)} \right) + \frac{\Delta G^{\ddagger}}{k_B T}\left(1 - \left(1 - \frac{\nu F_{eq} x^{\ddagger}}{\Delta G^{\ddagger}}\right)^{\frac{1}{\nu}}\right) \right) \right)^{\nu} \right)$$

Using the approximation from Eq. (A6) in Eq. (A7), we get Eq. (A8).

$$\langle F_{\text{Rev}} \rangle \approx \frac{\Delta G^{\ddagger}}{vx^{\ddagger}}\left(1-\left(1-\frac{k_{\text{B}}T}{\Delta G^{\ddagger}}\ln\left(e^{\frac{\Delta G^{\ddagger}}{k_{\text{B}}T}\left(1-\left(1-\frac{vF_{\text{eq}}x^{\ddagger}}{\Delta G^{\ddagger}}\right)^{\frac{1}{v}}\right)}+\frac{x^{\ddagger}\dot{F}e^{-\gamma}}{k_{0\text{un}}k_{\text{B}}T}\right)\right)^{v}\right) \qquad (A8)$$

Note that Eq. (A7) and Eq. (A8) are valid until the disappearance of the barrier. Mathematically, this is demonstrated by the negative values taken to the non-integer power. Next, we determine the minimum loading rate at which this occurs.

$$1-\frac{k_{\text{B}}T}{\Delta G^{\ddagger}}\ln\left(e^{\frac{\Delta G^{\ddagger}}{k_{\text{B}}T}\left(1-\left(1-\frac{vF_{\text{eq}}x^{\ddagger}}{\Delta G^{\ddagger}}\right)^{\frac{1}{v}}\right)}+\frac{x^{\ddagger}\dot{F}e^{-\gamma}}{k_{0}k_{\text{B}}T}\right)=0$$

$$e^{\frac{\Delta G^{\ddagger}}{k_{\text{B}}T}\left(1-\left(1-\frac{vF_{\text{eq}}x^{\ddagger}}{\Delta G^{\ddagger}}\right)^{\frac{1}{v}}\right)}+\frac{x^{\ddagger}\dot{F}e^{-\gamma}}{k_{0}k_{\text{B}}T}=e^{\frac{\Delta G^{\ddagger}}{k_{\text{B}}T}}$$

$$\dot{F}=\frac{k_{0}k_{\text{B}}Te^{\gamma}}{x^{\ddagger}}\left(e^{\frac{\Delta G^{\ddagger}}{k_{\text{B}}T}}-e^{\frac{\Delta G^{\ddagger}}{k_{\text{B}}T}\left(1-\left(1-\frac{vF_{\text{eq}}x^{\ddagger}}{\Delta G^{\ddagger}}\right)^{\frac{1}{v}}\right)}\right)$$

$$\dot{F}=\frac{k_{0}k_{\text{B}}Te^{\frac{\Delta G^{\ddagger}}{k_{\text{B}}T}+\gamma}}{x^{\ddagger}}\left(1-e^{-\frac{\Delta G^{\ddagger}}{k_{\text{B}}T}\left(1-\frac{vF_{\text{eq}}x^{\ddagger}}{\Delta G^{\ddagger}}\right)^{\frac{1}{v}}}\right)$$

At the critical loading rate, the mean reversible unfolding force is approximately $\Delta G^{\ddagger}/vx^{\ddagger}$ according to Eq. (A7) and Eq. (A8). After the barrier disappearance, the mean unfolding force increases as the square root of the loading rate $\sqrt{2\dot{F}x^{\ddagger}\zeta}$ [14,18]. In order to combine the two functions that are valid for the different loading rate regimes, we take the absolute value of the expression that is raised to the non-integer power in Eq. (A7) and use the Heaviside step function:

$$\langle F_{\mathrm{Rev}} \rangle \approx \frac{\Delta G^{\ddagger}}{vx^{\ddagger}} \left( 1 - \left| 1 - \frac{k_{\mathrm{B}}T}{\Delta G^{\ddagger}} \left( e^{\frac{k_0 k_{\mathrm{B}}T}{x^{\ddagger}\dot{F}} e^{\frac{\Delta G^{\ddagger}}{k_{\mathrm{B}}T}\left(1-\left(1-\frac{vF_{\mathrm{eq}}x^{\ddagger}}{\Delta G^{\ddagger}}\right)^{\frac{1}{v}}\right)}} E_1\left( \frac{k_0 k_{\mathrm{B}}T}{x^{\ddagger}\dot{F}} e^{\frac{\Delta G^{\ddagger}}{k_{\mathrm{B}}T}\left(1-\left(1-\frac{vF_{\mathrm{eq}}x^{\ddagger}}{\Delta G^{\ddagger}}\right)^{\frac{1}{v}}\right)} \right) + \frac{\Delta G^{\ddagger}}{k_{\mathrm{B}}T}\left(1-\left(1-\frac{vF_{\mathrm{eq}}x^{\ddagger}}{\Delta G^{\ddagger}}\right)^{\frac{1}{v}}\right) \right) \right|^{v} \right) \cdot$$

$$\theta\left(\dot{F}_{\mathrm{cRev}}-\dot{F}\right) + \left(\frac{\Delta G^{\ddagger}}{vx^{\ddagger}} - \sqrt{2\dot{F}_{\mathrm{cRev}} x^{\ddagger}\zeta} + \sqrt{2\dot{F} x^{\ddagger}\zeta}\right) \cdot \theta\left(\dot{F}-\dot{F}_{\mathrm{cRev}}\right)$$

(A9)

where $\dot{F}_{\mathrm{cRev}} = \frac{k_0 k_{\mathrm{B}} T e^{\frac{\Delta G^{\ddagger}}{k_{\mathrm{B}}T}+\gamma}}{x^{\ddagger}} \left(1 - e^{-\frac{\Delta G^{\ddagger}}{k_{\mathrm{B}}T}\left(1-\frac{vF_{\mathrm{eq}}x^{\ddagger}}{\Delta G^{\ddagger}}\right)^{\frac{1}{v}}}\right)$ and $k_0 = \frac{3\Delta G^{\ddagger}}{\pi \zeta x^{\ddagger 2}} e^{-\frac{\Delta G^{\ddagger}}{k_{\mathrm{B}}T}}$.

Eq. (A9) can be further simplified to Eq. (20), or using the approximation from Eq. (A6) to Eq. (A10).

$$\langle F_{\mathrm{Rev}} \rangle \approx \frac{\Delta G^{\ddagger}}{vx^{\ddagger}} \left( 1 - \left| 1 - \frac{k_{\mathrm{B}}T}{\Delta G^{\ddagger}} \ln\left( e^{\frac{\Delta G^{\ddagger}}{k_{\mathrm{B}}T}\left(1-\left(1-\frac{vF_{\mathrm{eq}}x^{\ddagger}}{\Delta G^{\ddagger}}\right)^{\frac{1}{v}}\right)} + \frac{e^{-\gamma}x^{\ddagger}\dot{F}}{k_{0\mathrm{un}}k_{\mathrm{B}}T} \right) \right|^{v} \right) \cdot \theta\left(\dot{F}_{\mathrm{cRev}}-\dot{F}\right) +$$

$$\left(\frac{\Delta G^{\ddagger}}{vx^{\ddagger}} - \sqrt{2\dot{F}_{\mathrm{cRev}} x^{\ddagger}\zeta} + \sqrt{2\dot{F} x^{\ddagger}\zeta}\right) \cdot \theta\left(\dot{F}-\dot{F}_{\mathrm{cRev}}\right)$$

(A10)

**Dependence of the mean first unfolding force on the loading rate for the unified model**

The dependence of the mean first unfolding force on the loading rate Eq. (19) can be obtained by starting the integration in Eq. (A3) from 0 and following the same steps, or by directly setting $F_{eq} = 0$ in Eq. (A9) and further simplifying.

**Dependence of the mean last unfolding force on the loading rate for the unified model**

The mean first refolding force Eq. (22) and the mean reversible refolding force Eq. (23) that are used for calculating the mean last unfolding force in Eq. (21) can be obtained using the same methods as described above, though by using Eq. (A11) for the refolding rate[41] and by using limits of integration for $\langle F_{\text{FirstRef}} \rangle$ from $F_{\text{fin}}$ to $F$ in Eq. (A13) and for $\langle F_{\text{RevRef}} \rangle$ from $F_{eq}$ to $F$ in a similar equation.

$$k_{\text{ref}}(F) = k_{0\text{ref}} \left(1 + \frac{vFx_{\text{ref}}^{\ddagger}}{\Delta G_{\text{ref}}^{\ddagger}}\right)^{\frac{1}{v}-1} e^{\frac{\Delta G_{\text{ref}}^{\ddagger}}{k_B T}\left[1-\left(1+\frac{vFx_{\text{ref}}^{\ddagger}}{\Delta G_{\text{ref}}^{\ddagger}}\right)^{\frac{1}{v}}\right]} \quad (A11)$$

$$\int k_{\text{ref}}(F) dF = -\frac{k_{0\text{ref}} k_B T}{x_{\text{ref}}^{\ddagger}} e^{\frac{\Delta G_{\text{ref}}^{\ddagger}}{k_B T}\left[1-\left(1+\frac{vFx_{\text{ref}}^{\ddagger}}{\Delta G_{\text{ref}}^{\ddagger}}\right)^{\frac{1}{v}}\right]} \quad (A12)$$

By substituting Eq. (A12) into Eq. (A13) we get Eq. (A14).

$$\int_1^p \frac{dp'}{p'} = \frac{1}{\dot{F}} \int_{F_{\text{fin}}}^F k_{\text{ref}}(F') dF' \quad (A13)$$

$$\ln p = -\frac{k_{0\text{ref}} k_B T}{x_{\text{ref}}^{\ddagger} \dot{F}} \left( e^{\frac{\Delta G_{\text{ref}}^{\ddagger}}{k_B T}\left[1-\left(1+\frac{vFx_{\text{ref}}^{\ddagger}}{\Delta G_{\text{ref}}^{\ddagger}}\right)^{\frac{1}{v}}\right]} - e^{\frac{\Delta G_{\text{ref}}^{\ddagger}}{k_B T}\left[1-\left(1+\frac{vF_{\text{fin}}x_{\text{ref}}^{\ddagger}}{\Delta G_{\text{ref}}^{\ddagger}}\right)^{\frac{1}{v}}\right]} \right) \quad (A14)$$

By solving Eq. (A14) for $F$ we find Eq. (A15).

$$F = \frac{\Delta G^{\ddagger}_{\text{ref}}}{v x^{\ddagger}_{\text{ref}}} \left( \left( 1 - \frac{k_B T}{\Delta G^{\ddagger}_{\text{ref}}} \ln \left( e^{\frac{\Delta G^{\ddagger}_{\text{ref}}}{k_B T} \left( 1 - \left( 1 + \frac{v F_{\text{fin}} x^{\ddagger}_{\text{ref}}}{\Delta G^{\ddagger}_{\text{ref}}} \right)^{\frac{1}{v}} \right)} - \frac{x^{\ddagger}_{\text{ref}} \dot{F} \ln p}{k_{0\text{ref}} k_B T} \right) \right)^{v} - 1 \right) \quad (A15)$$

At this step, we follow the approximate averaging procedure as in[22].

$$\alpha \equiv \ln \left( e^{\frac{\Delta G^{\ddagger}_{\text{ref}}}{k_B T} \left( 1 - \left( 1 + \frac{v F_{\text{fin}} x^{\ddagger}_{\text{ref}}}{\Delta G^{\ddagger}_{\text{ref}}} \right)^{\frac{1}{v}} \right)} - \frac{x^{\ddagger}_{\text{ref}} \dot{F} \ln p}{k_{0\text{ref}} k_B T} \right)$$

$$\langle F \rangle \approx F(\langle \alpha \rangle)$$

From Eq. (A15) and Eq. (A6), we obtain Eq. (A16) for the mean first refolding force.

$$\langle F_{\text{FirstRef}} \rangle \approx \frac{\Delta G^{\ddagger}_{\text{ref}}}{v x^{\ddagger}_{\text{ref}}} \left( \left( 1 - \frac{k_B T}{\Delta G^{\ddagger}_{\text{ref}}} \ln \left( e^{\frac{k_{0\text{ref}} k_B T}{x^{\ddagger}_{\text{ref}} \dot{F}} e^{\frac{\Delta G^{\ddagger}_{\text{ref}}}{k_B T} \left( 1 - \left( 1 + \frac{v F_{\text{fin}} x^{\ddagger}_{\text{ref}}}{\Delta G^{\ddagger}_{\text{ref}}} \right)^{\frac{1}{v}} \right)}} E_1 \left( \frac{k_{0\text{ref}} k_B T}{x^{\ddagger}_{\text{ref}} \dot{F}} e^{\frac{\Delta G^{\ddagger}_{\text{ref}}}{k_B T} \left( 1 - \left( 1 + \frac{v F_{\text{fin}} x^{\ddagger}_{\text{ref}}}{\Delta G^{\ddagger}_{\text{ref}}} \right)^{\frac{1}{v}} \right)} \right) + \frac{\Delta G^{\ddagger}_{\text{ref}}}{k_B T} \left( 1 - \left( 1 + \frac{v F_{\text{fin}} x^{\ddagger}_{\text{ref}}}{\Delta G^{\ddagger}_{\text{ref}}} \right)^{\frac{1}{v}} \right) \right) \right)^{v} - 1 \right)$$

(A16)

Using the approximation from Eq. (A6) in Eq. (A16), we get Eq. (A17).

$$\langle F_{\text{FirstRef}} \rangle \approx \frac{\Delta G^{\ddagger}_{\text{ref}}}{v x^{\ddagger}_{\text{ref}}} \left( \left( 1 - \frac{k_B T}{\Delta G^{\ddagger}_{\text{ref}}} \ln \left( e^{\frac{\Delta G^{\ddagger}_{\text{ref}}}{k_B T} \left( 1 - \left( 1 + \frac{v F_{\text{fin}} x^{\ddagger}_{\text{ref}}}{\Delta G^{\ddagger}_{\text{ref}}} \right)^{\frac{1}{v}} \right)} + \frac{e^{-\gamma} x^{\ddagger}_{\text{ref}} \dot{F}}{k_{0\text{ref}} k_B T} \right) \right)^{v} - 1 \right) \quad (A17)$$

Next, we find the critical loading rate in the same way as previously.

$$1 - \frac{k_B T}{\Delta G^{\ddagger}_{\text{ref}}} \ln \left( e^{\frac{\Delta G^{\ddagger}_{\text{ref}}}{k_B T} \left( 1 - \left( 1 + \frac{v F_{\text{fin}} x^{\ddagger}_{\text{ref}}}{\Delta G^{\ddagger}_{\text{ref}}} \right)^{\frac{1}{v}} \right)} + \frac{e^{-\gamma} x^{\ddagger}_{\text{ref}} \dot{F}}{k_{0\text{ref}} k_B T} \right) = 0$$

$$\dot{F} = \frac{k_{0\text{ref}} k_B T e^{\frac{\Delta G^{\ddagger}_{\text{ref}}}{k_B T}+\gamma}}{x^{\ddagger}_{\text{ref}}} \left(1 - e^{-\frac{\Delta G^{\ddagger}_{\text{ref}}}{k_B T}\left(1+\frac{\nu F_{\text{fin}} x^{\ddagger}_{\text{ref}}}{\Delta G^{\ddagger}_{\text{ref}}}\right)^{\frac{1}{\nu}}}\right)$$

By combining the two functions that are valid for the different loading rate regimes as previously, using the Heaviside step function we obtain

$$\langle F_{\text{FirstRef}} \rangle \approx \frac{\Delta G^{\ddagger}_{\text{ref}}}{\nu x^{\ddagger}_{\text{ref}}} \left( \left| 1 - \frac{k_B T}{\Delta G^{\ddagger}_{\text{ref}}} \left( e^{\frac{k_{0\text{ref}} k_B T}{x^{\ddagger}_{\text{ref}} \dot{F}} e^{\frac{\Delta G^{\ddagger}_{\text{ref}}}{k_B T}\left(1-\left(1+\frac{\nu F_{\text{fin}} x^{\ddagger}_{\text{ref}}}{\Delta G^{\ddagger}_{\text{ref}}}\right)^{\frac{1}{\nu}}\right)}} E_1\left(\frac{k_{0\text{ref}} k_B T}{x^{\ddagger}_{\text{ref}} \dot{F}} e^{\frac{\Delta G^{\ddagger}_{\text{ref}}}{k_B T}\left(1-\left(1+\frac{\nu F_{\text{fin}} x^{\ddagger}_{\text{ref}}}{\Delta G^{\ddagger}_{\text{ref}}}\right)^{\frac{1}{\nu}}\right)}\right) + \frac{\Delta G^{\ddagger}_{\text{ref}}}{k_B T} \left(1 - \left(1+\frac{\nu F_{\text{fin}} x^{\ddagger}_{\text{ref}}}{\Delta G^{\ddagger}_{\text{ref}}}\right)^{\frac{1}{\nu}}\right)\right) - 1 \right|^{\nu} \right) \cdot$$

$$\theta\left(\dot{F}_{\text{cFirstRef}}-\dot{F}\right) + \left(-\frac{\Delta G^{\ddagger}_{\text{ref}}}{\nu x^{\ddagger}_{\text{ref}}} + \sqrt{2\dot{F}_{\text{cFirstRef}} x^{\ddagger}_{\text{ref}} \zeta} - \sqrt{2\dot{F} x^{\ddagger}_{\text{ref}} \zeta}\right) \cdot \theta\left(\dot{F} - \dot{F}_{\text{cFirstRef}}\right)$$

(A18)

where $\dot{F}_{\text{cFirstRef}} = \dfrac{k_{0\text{ref}} k_B T e^{\frac{\Delta G^{\ddagger}_{\text{ref}}}{k_B T}+\gamma}}{x^{\ddagger}_{\text{ref}}} \left(1 - e^{-\frac{\Delta G^{\ddagger}_{\text{ref}}}{k_B T}\left(1+\frac{\nu F_{\text{fin}} x^{\ddagger}_{\text{ref}}}{\Delta G^{\ddagger}_{\text{ref}}}\right)^{\frac{1}{\nu}}}\right)$ and $k_{0\text{ref}} = \dfrac{3\Delta G^{\ddagger}_{\text{ref}}}{\pi \zeta x^{\ddagger\,2}_{\text{ref}}} e^{-\frac{\Delta G^{\ddagger}_{\text{ref}}}{k_B T}}$.

Eq. (A18) can be further simplified to Eq. (22), or by using the approximation from Eq. (A6) in a similar equation to Eq. (A10).

The dependence of the mean reversible refolding force on the loading rate Eq. (23) can be obtained by starting the integration in Eq. (A13) from $F_{\text{eq}}$ and following the same steps, or by directly setting $F_{\text{fin}}$ to $F_{\text{eq}}$ in Eq. (A18) and further simplifying.

# APPENDIX B

**Dependence of the mean reversible unfolding force on the loading rate for the unified model when the force is applied through the spring**

When the force is being applied through the spring, we need to use an appropriate expression for the force-dependent unfolding rate Eq. (B1)[34], which then can be plugged into Eq. (B2).

$$k(F) = k_0 \left( Z^2 - \frac{vFx^{\ddagger}Z}{\Delta G^{\ddagger}} \right)^{\frac{1}{v}-1} e^{\frac{\Delta G^{\ddagger}}{k_B T}\left(1 - \left(Z^2 - \frac{vFx^{\ddagger}Z}{\Delta G^{\ddagger}}\right)^{\frac{1}{v}}\right)} \tag{B1}$$

$$\int k(F) dF = \frac{k_0 k_B T}{x^{\ddagger} Z} e^{\frac{\Delta G^{\ddagger}}{k_B T}\left(1 - \left(Z^2 - \frac{vFx^{\ddagger}Z}{\Delta G^{\ddagger}}\right)^{\frac{1}{v}}\right)} \tag{B2}$$

Using the approach of[32], by substituting Eq. (B2) into Eq. (B3) we get Eq. (B4).

$$\int_1^p \frac{dp'}{p'} = -\frac{1}{\dot{F}} \int_{F_{eq}}^F k(F') dF' \tag{B3}$$

$$\ln p = -\frac{k_0 k_B T}{x^{\ddagger} Z \dot{F}} \left( e^{\frac{\Delta G^{\ddagger}}{k_B T}\left(1 - \left(Z^2 - \frac{vFx^{\ddagger}Z}{\Delta G^{\ddagger}}\right)^{\frac{1}{v}}\right)} - e^{\frac{\Delta G^{\ddagger}}{k_B T}\left(1 - \left(Z^2 - \frac{vF_{eq}x^{\ddagger}Z}{\Delta G^{\ddagger}}\right)^{\frac{1}{v}}\right)} \right) \tag{B4}$$

By solving Eq. (B4) for $F$, we find Eq. (B5).

$$F = \frac{\Delta G^{\ddagger}}{vx^{\ddagger}Z} \left( Z^2 - \left( 1 - \frac{k_B T}{\Delta G^{\ddagger}} \ln \left( e^{\frac{\Delta G^{\ddagger}}{k_B T}\left(1 - \left(Z^2 - \frac{vF_{eq}x^{\ddagger}Z}{\Delta G^{\ddagger}}\right)^{\frac{1}{v}}\right)} - \frac{x^{\ddagger} \dot{F} Z \ln p}{k_0 k_B T} \right) \right)^v \right) \tag{B5}$$

At this step, we follow the approximate averaging procedure as in[22].

$$\alpha \equiv \ln \left( e^{\frac{\Delta G^{\ddagger}}{k_B T}\left(1 - \left(Z^2 - \frac{vF_{eq}x^{\ddagger}Z}{\Delta G^{\ddagger}}\right)^{\frac{1}{v}}\right)} - \frac{x^{\ddagger} \dot{F} Z \ln p}{k_0 k_B T} \right)$$

$$\langle F \rangle \approx F(\langle \alpha \rangle)$$

From Eq. (B5) and Eq. (A6) we obtain the mean reversible unfolding force:

$$\langle F_{Rev} \rangle \approx \frac{\Delta G^{\ddagger}}{vx^{\ddagger}Z}\left(Z^2 - \left(1 - \frac{k_B T}{\Delta G^{\ddagger}}\left(\frac{k_0 k_B T}{e^{x^{\ddagger}\dot{F}Z}}e^{\frac{\Delta G^{\ddagger}}{k_B T}\left(1-\left(Z^2-\frac{vF_{eq}x^{\ddagger}Z}{\Delta G^{\ddagger}}\right)^{\frac{1}{v}}\right)}E_1\left(\frac{k_0 k_B T}{x^{\ddagger}\dot{F}Z}e^{\frac{\Delta G^{\ddagger}}{k_B T}\left(1-\left(Z^2-\frac{vF_{eq}x^{\ddagger}Z}{\Delta G^{\ddagger}}\right)^{\frac{1}{v}}\right)}+\frac{\Delta G^{\ddagger}}{k_B T}\left(1-\left(Z^2-\frac{vF_{eq}x^{\ddagger}Z}{\Delta G^{\ddagger}}\right)^{\frac{1}{v}}\right)\right)\right)\right)^v\right)$$

(B6)

Using the approximation from Eq. (A6) in Eq. (B6), we get Eq. (B7).

$$\langle F_{Rev} \rangle \approx \frac{\Delta G^{\ddagger}}{vx^{\ddagger}Z}\left(Z^2 - \left(1 - \frac{k_B T}{\Delta G^{\ddagger}}\ln\left(e^{\frac{\Delta G^{\ddagger}}{k_B T}\left(1-\left(Z^2-\frac{vF_{eq}x^{\ddagger}Z}{\Delta G^{\ddagger}}\right)^{\frac{1}{v}}\right)}+\frac{x^{\ddagger}\dot{F}Ze^{-\gamma}}{k_0 k_B T}\right)\right)^v\right) \qquad (B7)$$

Next, we find the critical loading rate in the same way as in Appendix A.

$$1 - \frac{k_B T}{\Delta G^{\ddagger}}\ln\left(e^{\frac{\Delta G^{\ddagger}}{k_B T}\left(1-\left(Z^2-\frac{vF_{eq}x^{\ddagger}Z}{\Delta G^{\ddagger}}\right)^{\frac{1}{v}}\right)}+\frac{x^{\ddagger}\dot{F}Ze^{-\gamma}}{k_0 k_B T}\right) = 0$$

$$\dot{F} = \frac{k_0 k_B T e^{\frac{\Delta G^{\ddagger}}{k_B T}+\gamma}}{x^{\ddagger}Z}\left(1 - e^{-\frac{\Delta G^{\ddagger}}{k_B T}\left(Z^2-\frac{vF_{eq}x^{\ddagger}Z}{\Delta G^{\ddagger}}\right)^{\frac{1}{v}}}\right)$$

By combining the two functions that are valid for the different loading rate regimes as was done in Appendix A, using the Heaviside step function we obtain

$$\langle F_{\text{Rev}} \rangle \approx \frac{\Delta G^{\ddagger}}{v x^{\ddagger} Z} \left( Z^2 - \left| 1 - \frac{k_B T}{\Delta G^{\ddagger}} \left[ \frac{k_0 k_B T}{x^{\ddagger} \dot{F} Z} e^{\frac{\Delta G^{\ddagger}}{k_B T}\left[1-\left(Z^2-\frac{v F_{\text{eq}} x^{\ddagger} Z}{\Delta G^{\ddagger}}\right)^{\frac{1}{v}}\right]} E_1 \left( \frac{k_0 k_B T}{x^{\ddagger} \dot{F} Z} e^{\frac{\Delta G^{\ddagger}}{k_B T}\left[1-\left(Z^2-\frac{v F_{\text{eq}} x^{\ddagger} Z}{\Delta G^{\ddagger}}\right)^{\frac{1}{v}}\right]} \right) + \frac{\Delta G^{\ddagger}}{k_B T} \left( 1 - \left( Z^2 - \frac{v F_{\text{eq}} x^{\ddagger} Z}{\Delta G^{\ddagger}} \right)^{\frac{1}{v}} \right) \right]^{v} \right) \cdot$$

$$\theta\left(\dot{F}_{\text{cRev}} - \dot{F}\right) + \left( \frac{\Delta G^{\ddagger} Z}{v x^{\ddagger}} - \sqrt{2 \dot{F}_{\text{cRev}} x^{\ddagger} \zeta} + \sqrt{2 \dot{F} x^{\ddagger} \zeta} \right) \cdot \theta\left(\dot{F} - \dot{F}_{\text{cRev}}\right)$$

(B8)

$$\text{where } \dot{F}_{\text{cRev}} = \frac{k_0 k_B T e^{\frac{\Delta G^{\ddagger}}{k_B T} + \gamma}}{x^{\ddagger} Z} \left( 1 - e^{-\frac{\Delta G^{\ddagger}}{k_B T}\left(Z^2 - \frac{v F_{\text{eq}} x^{\ddagger} Z}{\Delta G^{\ddagger}}\right)^{\frac{1}{v}}} \right)$$

Eq. (B8) can be further simplified to Eq. (25), or by using the approximation from Eq. (A6) to a similar equation to Eq. (A10).

**Dependence of the mean first unfolding force on the loading rate for the unified model when the force is applied through the spring**

The dependence of the mean first unfolding force on the loading rate Eq. (24) can be obtained by starting the integration in Eq. (B3) from 0 and following the same steps, or by directly setting $F_{\text{eq}} = 0$ in Eq. (B8) and further simplifying.

**Dependence of the mean last unfolding force on the loading rate for the unified model when the force is applied through the spring**

The mean first refolding force Eq. (26) and the mean reversible refolding force Eq. (27) that are used for calculating the mean last unfolding force when the force is applied through the spring in Eq. (21) can be obtained using the same methods as described above, using Eq. (B9) for the refolding rate[41] and limits of integration for $\langle F_{\text{FirstRef}} \rangle$ from $F_{\text{fin}}$ to $F$ in Eq. (B11) and for $\langle F_{\text{RevRef}} \rangle$ from $F_{\text{eq}}$ to $F$ in a similar equation.

$$k_{\text{ref}}(F) = k_{0\text{ref}} \left( Z_{\text{ref}}^2 + \frac{\nu F x_{\text{ref}}^{\ddagger} Z_{\text{ref}}}{\Delta G_{\text{ref}}^{\ddagger}} \right)^{\frac{1}{\nu}-1} e^{\frac{\Delta G_{\text{ref}}^{\ddagger}}{k_B T} \left( 1 - \left( Z_{\text{ref}}^2 + \frac{\nu F x_{\text{ref}}^{\ddagger} Z_{\text{ref}}}{\Delta G_{\text{ref}}^{\ddagger}} \right)^{\frac{1}{\nu}} \right)} \tag{B9}$$

$$\int k_{\text{ref}}(F) dF = -\frac{k_{0\text{ref}} k_B T}{x_{\text{ref}}^{\ddagger} Z_{\text{ref}}} e^{\frac{\Delta G_{\text{ref}}^{\ddagger}}{k_B T} \left( 1 - \left( Z_{\text{ref}}^2 + \frac{\nu F x_{\text{ref}}^{\ddagger} Z_{\text{ref}}}{\Delta G_{\text{ref}}^{\ddagger}} \right)^{\frac{1}{\nu}} \right)} \tag{B10}$$

By substituting Eq. (B10) into Eq. (B11) we get Eq. (B12).

$$\int_1^p \frac{dp'}{p'} = \frac{1}{\dot{F}} \int_{F_{\text{fin}}}^F k_{\text{ref}}(F') dF' \tag{B11}$$

$$\ln p = -\frac{k_{0\text{ref}} k_B T}{x_{\text{ref}}^{\ddagger} \dot{F} Z_{\text{ref}}} \left( e^{\frac{\Delta G_{\text{ref}}^{\ddagger}}{k_B T} \left( 1 - \left( Z_{\text{ref}}^2 + \frac{\nu F x_{\text{ref}}^{\ddagger} Z_{\text{ref}}}{\Delta G_{\text{ref}}^{\ddagger}} \right)^{\frac{1}{\nu}} \right)} - e^{\frac{\Delta G_{\text{ref}}^{\ddagger}}{k_B T} \left( 1 - \left( Z_{\text{ref}}^2 + \frac{\nu F_{\text{fin}} x_{\text{ref}}^{\ddagger} Z_{\text{ref}}}{\Delta G_{\text{ref}}^{\ddagger}} \right)^{\frac{1}{\nu}} \right)} \right) \tag{B12}$$

By solving Eq. (B12) for $F$, we find Eq. (B13).

$$F = \frac{\Delta G_{\text{ref}}^{\ddagger}}{\nu x_{\text{ref}}^{\ddagger} Z_{\text{ref}}} \left( \left( 1 - \frac{k_B T}{\Delta G_{\text{ref}}^{\ddagger}} \ln \left( e^{\frac{\Delta G_{\text{ref}}^{\ddagger}}{k_B T} \left( 1 - \left( Z_{\text{ref}}^2 + \frac{\nu F_{\text{fin}} x_{\text{ref}}^{\ddagger} Z_{\text{ref}}}{\Delta G_{\text{ref}}^{\ddagger}} \right)^{\frac{1}{\nu}} \right)} - \frac{x_{\text{ref}}^{\ddagger} \dot{F} Z_{\text{ref}} \ln p}{k_{0\text{ref}} k_B T} \right) \right)^{\nu} - Z_{\text{ref}}^2 \right) \tag{B13}$$

At this step, we follow the approximate averaging procedure as in[22].

$$\alpha \equiv \ln \left( e^{\frac{\Delta G_{\text{ref}}^{\ddagger}}{k_B T} \left( 1 - \left( Z_{\text{ref}}^2 + \frac{\nu F_{\text{fin}} x_{\text{ref}}^{\ddagger} Z_{\text{ref}}}{\Delta G_{\text{ref}}^{\ddagger}} \right)^{\frac{1}{\nu}} \right)} - \frac{x_{\text{ref}}^{\ddagger} \dot{F} Z_{\text{ref}} \ln p}{k_{0\text{ref}} k_B T} \right)$$

$$\langle F \rangle \approx F(\langle \alpha \rangle)$$

From Eq. (B13) and Eq. (A6), we obtain for the mean first refolding force Eq. (B14).

$$\langle F_{\text{FirstRef}} \rangle \approx \frac{\Delta G^{\ddagger}_{\text{ref}}}{v x^{\ddagger}_{\text{ref}} Z_{\text{ref}}} \left( \left( 1 - \frac{k_B T}{\Delta G^{\ddagger}_{\text{ref}}} e^{\frac{k_{0\text{ref}} k_B T}{x^{\ddagger}_{\text{ref}} \dot{F} Z_{\text{ref}}} e^{\frac{\Delta G^{\ddagger}_{\text{ref}}}{k_B T} \left( 1 - \left( Z^2_{\text{ref}} + \frac{v F_{\text{fin}} x^{\ddagger}_{\text{ref}} Z_{\text{ref}}}{\Delta G^{\ddagger}_{\text{ref}}} \right)^{\frac{1}{v}} \right)}} \left( E_1 \left( \frac{k_{0\text{ref}} k_B T}{x^{\ddagger}_{\text{ref}} \dot{F} Z_{\text{ref}}} e^{\frac{\Delta G^{\ddagger}_{\text{ref}}}{k_B T} \left( 1 - \left( Z^2_{\text{ref}} + \frac{v F_{\text{fin}} x^{\ddagger}_{\text{ref}} Z_{\text{ref}}}{\Delta G^{\ddagger}_{\text{ref}}} \right)^{\frac{1}{v}} \right)}} \right) + \frac{\Delta G^{\ddagger}_{\text{ref}}}{k_B T} \left( 1 - \left( Z^2_{\text{ref}} + \frac{v F_{\text{fin}} x^{\ddagger}_{\text{ref}} Z_{\text{ref}}}{\Delta G^{\ddagger}_{\text{ref}}} \right)^{\frac{1}{v}} \right) \right) \right)^{v} - Z^2_{\text{ref}} \right)$$

(B14)

Using the approximation from Eq. (A6) in Eq. (B14), we get Eq. (B15).

$$\langle F_{\text{FirstRef}} \rangle \approx \frac{\Delta G^{\ddagger}_{\text{ref}}}{v x^{\ddagger}_{\text{ref}} Z_{\text{ref}}} \left( \left( 1 - \frac{k_B T}{\Delta G^{\ddagger}_{\text{ref}}} \ln \left( e^{\frac{\Delta G^{\ddagger}_{\text{ref}}}{k_B T} \left( 1 - \left( Z^2_{\text{ref}} + \frac{v F_{\text{fin}} x^{\ddagger}_{\text{ref}} Z_{\text{ref}}}{\Delta G^{\ddagger}_{\text{ref}}} \right)^{\frac{1}{v}} \right)} + \frac{x^{\ddagger}_{\text{ref}} \dot{F} Z_{\text{ref}} e^{-\gamma}}{k_{0\text{ref}} k_B T} \right) \right)^{v} - Z^2_{\text{ref}} \right) \quad \text{(B15)}$$

Next, we find the critical loading rate in the same way as previously.

$$1 - \frac{k_B T}{\Delta G^{\ddagger}_{\text{ref}}} \ln \left( e^{\frac{\Delta G^{\ddagger}_{\text{ref}}}{k_B T} \left( 1 - \left( Z^2_{\text{ref}} + \frac{v F_{\text{fin}} x^{\ddagger}_{\text{ref}} Z_{\text{ref}}}{\Delta G^{\ddagger}_{\text{ref}}} \right)^{\frac{1}{v}} \right)} + \frac{x^{\ddagger}_{\text{ref}} \dot{F} Z_{\text{ref}} e^{-\gamma}}{k_{0\text{ref}} k_B T} \right) = 0$$

$$\dot{F} = \frac{k_{0\text{ref}} k_B T e^{\frac{\Delta G^{\ddagger}_{\text{ref}}}{k_B T} + \gamma}}{x^{\ddagger}_{\text{ref}} Z_{\text{ref}}} \left( 1 - e^{-\frac{\Delta G^{\ddagger}_{\text{ref}}}{k_B T} \left( Z^2_{\text{ref}} + \frac{v F_{\text{fin}} x^{\ddagger}_{\text{ref}} Z_{\text{ref}}}{\Delta G^{\ddagger}_{\text{ref}}} \right)^{\frac{1}{v}}} \right)$$

By combining the two functions that are valid for the different loading rate regimes as previously, using the Heaviside step function we obtain Eq. (B16):

$$\langle F_{\text{FirstRef}} \rangle \approx \frac{\Delta G^{\ddagger}_{\text{ref}}}{v x^{\ddagger}_{\text{ref}} Z_{\text{ref}}} \left( \left( 1 - \frac{k_B T}{\Delta G^{\ddagger}_{\text{ref}}} e^{\frac{k_{0\text{ref}} k_B T}{x^{\ddagger}_{\text{ref}} \dot{F} Z_{\text{ref}}} e^{\frac{\Delta G^{\ddagger}_{\text{ref}}}{k_B T} \left( 1 - \left( Z^2_{\text{ref}} + \frac{v F_{\text{fin}} x^{\ddagger}_{\text{ref}} Z_{\text{ref}}}{\Delta G^{\ddagger}_{\text{ref}}} \right)^{\frac{1}{v}} \right)}} \left( E_1 \left( \frac{k_{0\text{ref}} k_B T}{x^{\ddagger}_{\text{ref}} \dot{F} Z_{\text{ref}}} e^{\frac{\Delta G^{\ddagger}_{\text{ref}}}{k_B T} \left( 1 - \left( Z^2_{\text{ref}} + \frac{v F_{\text{fin}} x^{\ddagger}_{\text{ref}} Z_{\text{ref}}}{\Delta G^{\ddagger}_{\text{ref}}} \right)^{\frac{1}{v}} \right)}} \right) + \frac{\Delta G^{\ddagger}_{\text{ref}}}{k_B T} \left( 1 - \left( Z^2_{\text{ref}} + \frac{v F_{\text{fin}} x^{\ddagger}_{\text{ref}} Z_{\text{ref}}}{\Delta G^{\ddagger}_{\text{ref}}} \right)^{\frac{1}{v}} \right) \right) \right)^{v} - Z^2_{\text{ref}} \right) \cdot$$

$$\theta \left( \dot{F}_{\text{cFirst}} - \dot{F} \right) + \left( -\frac{\Delta G^{\ddagger}_{\text{ref}} Z_{\text{ref}}}{v x^{\ddagger}_{\text{ref}}} + \sqrt{2 \dot{F}_{\text{cFirst}} x^{\ddagger}_{\text{ref}} \zeta} - \sqrt{2 \dot{F} x^{\ddagger}_{\text{ref}} \zeta} \right) \cdot \theta \left( \dot{F} - \dot{F}_{\text{cFirst}} \right)$$

(B16)

where $\dot{F}_{\text{cFirst}} = \dfrac{k_{\text{0ref}} k_{\text{B}} T e^{\frac{\Delta G^{\ddagger}_{\text{ref}}}{k_{\text{B}} T} + \gamma}}{x^{\ddagger}_{\text{ref}} Z_{\text{ref}}} \left( 1 - e^{-\frac{\Delta G^{\ddagger}_{\text{ref}}}{k_{\text{B}} T} \left( Z^2_{\text{ref}} + \frac{\nu F_{\text{fin}} x^{\ddagger}_{\text{ref}} Z_{\text{ref}}}{\Delta G^{\ddagger}_{\text{ref}}} \right)^{\frac{1}{\nu}}} \right)$

Eq. (B16) can be further simplified to Eq. (26) or using the approximation from Eq. (A6) to a similar equation to Eq. (A10).

The dependence of the mean reversible refolding force on the loading rate when the force is applied through a spring, given by Eq. (27), can be obtained by starting the integration in Eq. (B11) from $F_{\text{eq}}$ and following the same steps, or by directly setting $F_{\text{fin}}$ to $F_{\text{eq}}$ in Eq. (B16) and further simplifying.